\newcommand\apjcls{1}
\newcommand\aastexcls{2}
\newcommand\othercls{3}

\newcommand\papercls{\aastexcls}
\documentclass[tighten, times,twocolumn]{aastex62}

\if\papercls \apjcls
\usepackage{apjfonts}
\else\if\papercls \othercls
\usepackage{epsfig}
\usepackage{margin}
\usepackage{times}
\fi\fi
\usepackage{ifthen}
\usepackage{natbib}
\usepackage{bm}
\usepackage{amssymb, amsmath}
\usepackage{appendix}
\usepackage{etoolbox}
\usepackage[T1]{fontenc}
\usepackage{paralist}
\usepackage{newtxtext,newtxmath}
\if\papercls \apjcls 
\newcommand\aas{\ref@jnl{AAS Meeting Abstracts}}
\newcommand\dps{\ref@jnl{AAS/DPS Meeting Abstracts}}
\newcommand\maps{\ref@jnl{MAPS}}
\else\if\papercls \othercls
\usepackage{astjnlabbrev-jh}
\fi\fi

\if\papercls \aastexcls
\hypersetup{citecolor=blue, 
            linkcolor=blue, 
            menucolor=blue, 
            urlcolor=blue}  
\else
\usepackage[
bookmarks=true,           
bookmarksnumbered=true,   
colorlinks=true,          
citecolor=blue,           
linkcolor=blue,           
menucolor=blue,           
urlcolor=blue,            
linkbordercolor={0 0 1},  
pdfborder={0 0 1},
frenchlinks=true]{hyperref}
\fi

\if\papercls \othercls

\else

\fi

\providecommand{\adsurl}[1]{\href{#1}{ADS}}

\makeatletter
\patchcmd{\NAT@citex}
  {\@citea\NAT@hyper@{%
     \NAT@nmfmt{\NAT@nm}%
     \hyper@natlinkbreak{\NAT@aysep\NAT@spacechar}{\@citeb\@extra@b@citeb}%
     \NAT@date}}
  {\@citea\NAT@nmfmt{\NAT@nm}%
   \NAT@aysep\NAT@spacechar\NAT@hyper@{\NAT@date}}{}{}

\patchcmd{\NAT@citex}
  {\@citea\NAT@hyper@{%
     \NAT@nmfmt{\NAT@nm}%
     \hyper@natlinkbreak{\NAT@spacechar\NAT@@open\if*#1*\else#1\NAT@spacechar\fi}%
       {\@citeb\@extra@b@citeb}%
     \NAT@date}}
  {\@citea\NAT@nmfmt{\NAT@nm}%
   \NAT@spacechar\NAT@@open\if*#1*\else#1\NAT@spacechar\fi\NAT@hyper@{\NAT@date}}
\makeatother

\makeatletter
\DeclareRobustCommand{\lowcase}[1]{\@lowcase#1\@nil}
\def\@lowcase#1\@nil{\if\relax#1\relax\else\MakeLowercase{#1}\fi}
\makeatother

\DeclareSymbolFont{UPM}{U}{eur}{m}{n}
\DeclareMathSymbol{\umu}{0}{UPM}{"16}
\let\oldumu=\umu
\renewcommand\umu{\ifmmode\oldumu\else\math{\oldumu}\fi}

\if\papercls \othercls

\else

\fi

\let\oldsim=\sim
\renewcommand\sim{\ifmmode\oldsim\else\math{\oldsim}\fi}
\let\oldpm=\pm
\renewcommand\pm{\ifmmode\oldpm\else\math{\oldpm}\fi}
\newcommand\by{\ifmmode\times\else\math{\times}\fi}

\newbox{\wdbox}
\renewcommand\c{\setbox\wdbox=\hbox{,}\hspace{\wd\wdbox}}
\renewcommand\i{\setbox\wdbox=\hbox{i}\hspace{\wd\wdbox}}

\newcount\timect
\newcount\hourct
\newcount\minct
\newcommand\now{\timect=\time \divide\timect by 60
         \hourct=\timect Cltiply\hourct by 60
         \minct=\time \advance\minct by -\hourct
         \number\timect:\ifnum \minct < 10 0\fi\number\minct}

\catcode`@=11

\newcommand\comment[1]{}

\newcommand\commenton{\catcode`\%=14}

\renewcommand\math[1]{$#1$}
\newcommand\mathshifton{\catcode`\$=3}

\let\atab=&
\newcommand\atabon{\catcode`\&=4}

\let\oldmsp=\sp
\let\oldmsb=\sb
\def\sp#1{\ifmmode
           \oldmsp{#1}%
         \else\strut\raise.85ex\hbox{\scriptsize #1}\fi}
\def\sb#1{\ifmmode
           \oldmsb{#1}%
         \else\strut\raise-.54ex\hbox{\scriptsize #1}\fi}
\newbox\@sp
\newbox\@sb
\def\sbp#1#2{\ifmmode%
           \oldmsb{#1}\oldmsp{#2}%
         \else
           \setbox\@sb=\hbox{\sb{#1}}%
           \setbox\@sp=\hbox{\sp{#2}}%
           \rlap{\copy\@sb}\copy\@sp
           \ifdim \wd\@sb >\wd\@sp
             \hskip -\wd\@sp \hskip \wd\@sb
           \fi
        \fi}
\def\msp#1{\ifmmode
           \oldmsp{#1}
         \else \math{\oldmsp{#1}}\fi}
\def\msb#1{\ifmmode
           \oldmsb{#1}
         \else \math{\oldmsb{#1}}\fi}

\def\supon{\catcode`\^=7}

\def\subon{\catcode`\_=8}

\def\supsubon{\supon \subon}

\newcommand\actcharon{\catcode`\~=13}

\newcommand\paramon{\catcode`\#=6}

\comment{And now to turn us totally on and off...}

\newcommand\reservedcharson{ \commenton  \mathshifton  \atabon  \supsubon 
                             \actcharon  \paramon}

\catcode`@=12
\reservedcharson

\if\papercls \apjcls

\else

\fi

\newcommand\chisq{\ifmmode{\chi\sp{2}}\else\math{\chi\sp{2}}\fi}
\newcommand\redchisq{\ifmmode{ \chi\sp{2}\sb{\rm red}}
                    \else\math{\chi\sp{2}\sb{\rm red}}\fi}
\newcommand\Teq{\ifmmode{T\sb{\rm eq}}\else$T$\sb{eq}\fi}
\newcommand\mjup{\ifmmode{M\sb{\rm Jup}}\else$M$\sb{Jup}\fi}
\newcommand\rjup{\ifmmode{R\sb{\rm Jup}}\else$R$\sb{Jup}\fi}
\newcommand\msun{\ifmmode{M\sb{\odot}}\else$M\sb{\odot}$\fi}
\newcommand\rsun{\ifmmode{R\sb{\odot}}\else$R\sb{\odot}$\fi}
\newcommand\mearth{\ifmmode{M\sb{\oplus}}\else$M\sb{\oplus}$\fi}
\newcommand\rearth{\ifmmode{R\sb{\oplus}}\else$R\sb{\oplus}$\fi}

\shorttitle{Three-component Phase Separation in UMWDs}
\shortauthors{M. Castro-Tapia \& A. Cumming}

\begin{document}

\title{Three-component Phase Separation for Ultramassive White Dwarf Models}

\author{Matias Castro-Tapia}
\affiliation{\rm Department of Physics and Trottier Space Institute, McGill University, Montreal, QC H3A 2T8, Canada}
\email{matias.castrotapia@mail.mcgill.ca}

\author{Andrew Cumming}
\affiliation{\rm Department of Physics and Trottier Space Institute, McGill University, Montreal, QC H3A 2T8, Canada}


\begin{abstract}
We investigate phase separation in oxygen-neon (O/Ne) ultramassive white dwarfs (UMWDs). Current stellar evolution codes, such as MESA, only account for $\mathrm{^{16}O/^{20}Ne}$ separation and do not include other minor species. To improve this, we implement ternary phase diagrams into MESA. We construct UMWD models with O/Ne/sodium (Na) and O/Ne/magnesium (Mg) cores to test our implementation. We also assess the effect of including $\mathrm{^{22}Ne}$ in the current two-species framework. Our results show that incorporating additional components into the phase separation significantly alters the chemical evolution of UMWDs. Heavier elements preferentially enrich the solid core, enhancing mixing in the overlying liquid. We compute the buoyancy flux driven by compositional instabilities during crystallization. As in previous studies, we find two convective regimes: an early, fast overturning convection, lasting less than a million years, followed by inefficient (thermohaline) convection. The fast convective regime lasts up to 100 times longer with three-species separation compared to the standard $\mathrm{^{16}O/^{20}Ne}$ case. We find that neutron-rich species can have a significant contribution to the buoyancy flux despite their small mass fraction (<10\%). We compute the amount of cooling delay induced by phase separation in UMWDs, and find that the three-species phase separation produces a delay up to $\sim10$ times larger than the simplest case of fractionation, although still less than 1 Gyr. We predict that the change in the composition profile in the liquid region when three components are included should change the frequency of gravity modes that can propagate in the interior of pulsating UMWDs.
\end{abstract}


\section{Introduction}
Most single white dwarfs (WDs) are expected to form with an internal composition of carbon-oxygen (C/O). However, single WDs with masses $\gtrsim1.05\ M_{\odot}$ are believed to form oxygen-neon (O/Ne) cores as their progenitors in the asymptotic giant branch (AGB) can reach the conditions to trigger C-burning \citep{Siess2006}. While in typical WDs the internal composition of C/O is expected to account for $\gtrsim90\%$ \citep[e.g.][]{Bauer2023} of the core mixture, in ultramassive (UM) WDs ($\gtrsim1.05\ M_{\odot}$), the O/Ne mixture is expected to encompass a smaller fraction of the core composition $\sim 85\%-90\%$. Thus, significant amounts of C, sodium (Na), and magnesium (Mg) are predicted to be $\lesssim 15\%$ of the UMWD core composition \citep[e.g.][]{Camisassa2019, Bauer2020, Schwab2021, Blouin2021c}.


WDs cool over billions of years, and due to their extremely dense interiors, they experience a first-order phase transition when reaching a critical temperature \citep{vanHorn1968}. Given their interior density profile, they crystallize from the core outwards. Studies of phase diagrams of C/O and O/Ne mixtures applied to WDs predict that, as the crystal forms, it will be enriched in heavier elements compared to the original mixture \citep{SegretainChabrier1993, Horowitz2010, MedinCumming2010, Blouin2020, Blouin2021b, Blouin2021c}. Current WD cooling models include the implementation of this 2-species fractionation of the core chemical abundances as it crystallizes (e.g., \citealt{Althaus2012} for the LPCODE, \citealt{Salaris2022} for BaSTI, and \citealt{Bauer2023} for MESA). Such an implementation only accounts for the two most abundant isotopes in each mixture $^{12}\mathrm{C}$/$^{16}\mathrm{O}$ or $^{16}\mathrm{O}$/$^{20}\mathrm{Ne}$. To date, there are no publicly available models that include the crystallization of additional species, such as those that appear in UMWDs.

As the WD crystallization advances and the solid phase is enriched in heavy elements, the liquid phase on top of the core is depleted of heavy elements as mass and number of particles are conserved. This creates a buoyant, unstable region between the liquid layer enriched in light elements and the outer layer that did not fractionate \citep{Stevenson1980, Mochkovitch1983, Isern1997, Fuentes2023}, inducing convection that acts to mix the liquid phase. This element redistribution provides a source of extra energy that delays the cooling \citep[e.g.][]{Isern2000, Althaus2012, Blouin2020, Bauer2023}, which is key for explaining current observations \citep{Tremblay2019}. 
Furthermore, this fluid motion, if vigorous enough, can potentially power a dynamo, inducing a magnetic field in the WD interior \citep{Isern2017, Ginzburg2022, Fuentes2024}.

Recently, \citet{Castro-Tapia2024}, based on \citet{Fuentes2023}, demonstrated that two regimes of convection may be induced by WD crystallization, making plausible a dynamo only for a few Myr following crystallization. However, this study only focused on C/O WDs with masses $\leq 1\ M_{\odot}$ with the current 2-species phase separation in the stellar evolution code MESA. As single-evolved UMWDs with O/Ne cores contain significant amounts of other elements, in this paper, we use a 3-species phase diagram to study the case. We also include neutron-rich species with a ratio of atomic number to mass number $Z/A<1/2$, such as the $\mathrm{^{23}Na}$ in O/Ne WDs \citep[e.g.][]{Camisassa2019}, since they could raise the convective efficiency at the solid-liquid boundary when fractionated along with elements with $Z/A=1/2$, as in the case of crystallization in the ocean of accreting neutron stars \citep[e.g.][]{MedinCumming2015, Fuentes2023}. A similar effect plays a role for $\mathrm{^{22}Ne}$ (with $Z/A<1/2$) gravitational settling in the liquid phase for C/O WD interior \citep{BildstenHall2001}.

While some studies have already investigated the 3-component phase separation in WDs, those works have mostly focused on the delay of the WD cooling due to sedimentation of Ne-clusters (or Na-clusters, Fe-clusters) \citep{Bauer2020, Caplan2020, Caplan2021} and distillation of Ne-poor (or Mg-poor) crystals \citep{Isern1991, Blouin2021, Blouin2021c, Shen2023, Bedard2024}. While sedimentation has been found unlikely by molecular dynamics studies \citep{Caplan2020, Caplan2021}, the distillation theory has appeared as the most plausible scenario for explaining an overdensity of delayed massive WDs in the color-magnitude diagram \citep{Cheng2019}. However, the WDs that can drive distillation are most likely to be formed in C/O + Helium (He) WD mergers \citep{Wu2022} with C/O cores and impurities of $^{22}\mathrm{Ne}$; or subgiant + C/O WD mergers which also leads to $^{25}\mathrm{Mg}$, and $^{26}\mathrm{Mg}$ impurities within the C/O core \citep{Shen2023}.

 The already implemented 2-species phase separation in stellar evolution codes predicts a crystallization delay of about $\sim0.6$ Gyr for C/O cores \citep{Bauer2023}, and only $\sim0.1$ Gyr for O/Ne cores \citep{Camisassa2019}. On the other hand, \citet{Blouin2021c} investigated whether distillation could be driven by $^{23}\mathrm{Na}$ impurities in WDs with O/Ne cores, and found that the solid phase is always denser than the liquid, preventing distillation. Additionally, they concluded that impurities of $\mathrm{^{23}Ne}$ and $\mathrm{^{24}Mg}$ do not significantly affect the phase diagram given their low abundance ($X\lesssim10\%$). They then conclude that 2-species diagrams are safe for O/Ne WD models. However, the cooling delay these impurities can induce has not been quantified.

Another intriguing puzzle is the poorly-constrained internal composition of UMWDs. Although O/Ne-rich cores are usually predicted from classic single evolution models, some modifications during the AGB evolution \citep{Althaus2012}, or merger events can still produce C/O cores \citep[e.g.][]{Cheng2020}, and even high overshoot during the AGB can lead to unusual $\mathrm{^{12}C}$/$\mathrm{^{16}O}$/$\mathrm{^{20}Ne}$ hybrid cores \citep{SchwabGaraud2019}.  Asteroseismology offers a promising way to constrain the composition of these objects. For example, some UMWDs exhibit detectable photometric variations consistent with pulsations (specifically, $g$-modes), such as the ZZ Ceti or DAV WDs \citep[e.g.,][and references therein]{Kilic2023, DeGeronimo2025}. From these pulsations, many stellar parameters can be inferred \citep[e.g.][]{Calcaferro2024}. However, as many of these pulsating UMWDs are predicted to have a significant portion of their core crystallized \citep[e.g., more than $80\%$ for the cases studied by][]{Corsico2019}, accurate models for the phase separation are needed to improve predictions from asteroseismic measurements.

Although 3-component phase diagrams have been computed \citep{MedinCumming2010, Caplan2020, Blouin2021, Blouin2021c}, they have not yet been implemented in stellar evolution codes for models of O/Ne UMWDs. In this paper, we implement 3-component phase diagrams in MESA and use them to construct models of cooling UMWDs. In Section \ref{sec:phase_sep}, we present the 3-component phase diagrams constructed and the methods used to modify the current routine in MESA to implement the fractionation of 3 species. Additionally, we also address the lack of inclusion of $^{22}\mathrm{Ne}$ in the current version of the O/Ne phase separation in MESA, as some amount of this isotope is also present in UMWD interiors and will separate together with the $^{20}\mathrm{Ne}$. In 
Section \ref{sec:WD_models}, we construct UMWD models to test our modified routine with 3 species. We then compare to models that include only 2-species phase diagrams. W  e show three possible applications to improve calculations in the context of crystallization-driven convection, cooling curves, and models for pulsating WDs. Finally, in Section \ref{sec:sum_disc}, we summarize our findings and discuss future directions.

\begin{figure*}
    \centering
    \includegraphics[width=0.32\textwidth]{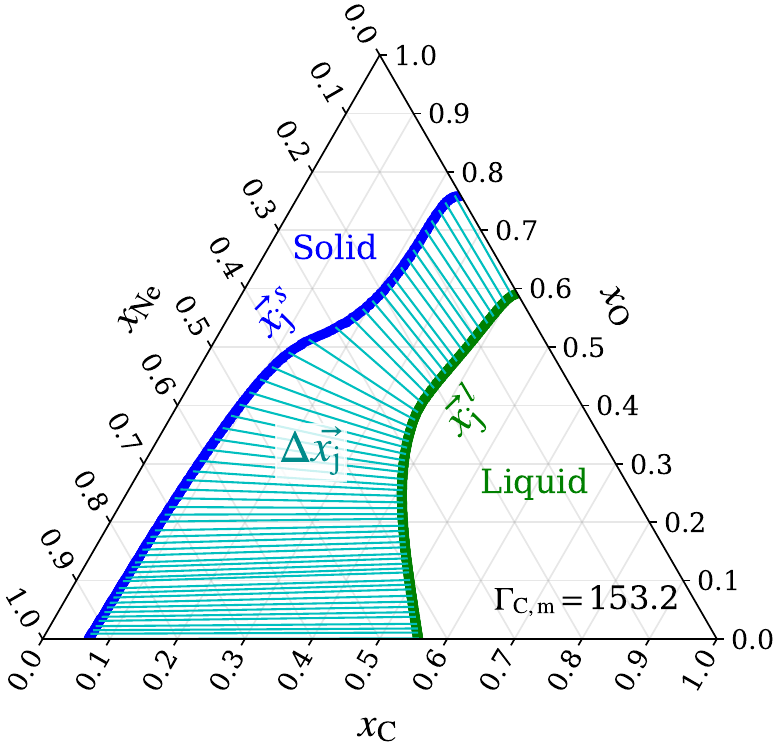}\hspace{0.in}
    \includegraphics[width=0.32\textwidth]{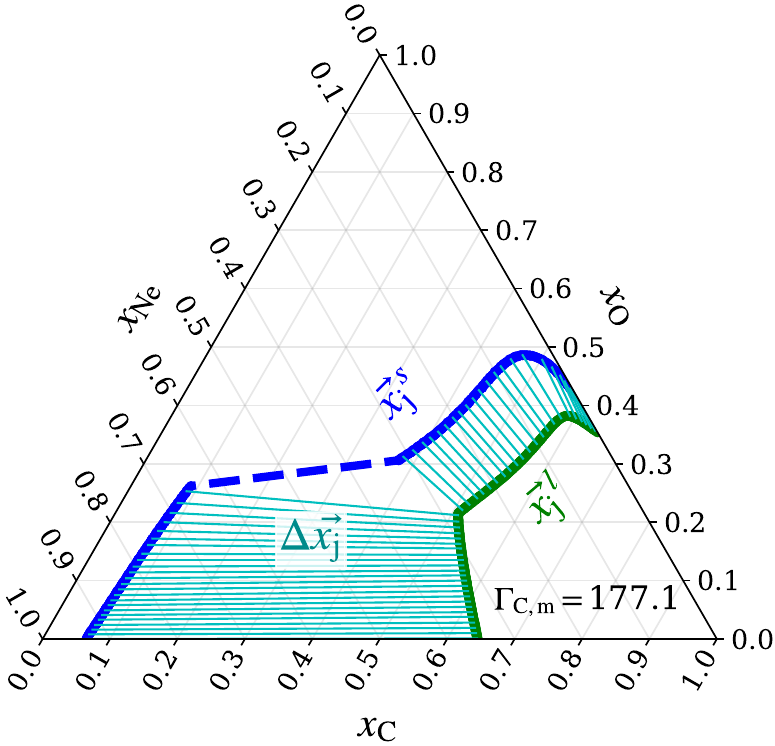}\hspace{0.in}
    \includegraphics[width=0.33\textwidth]{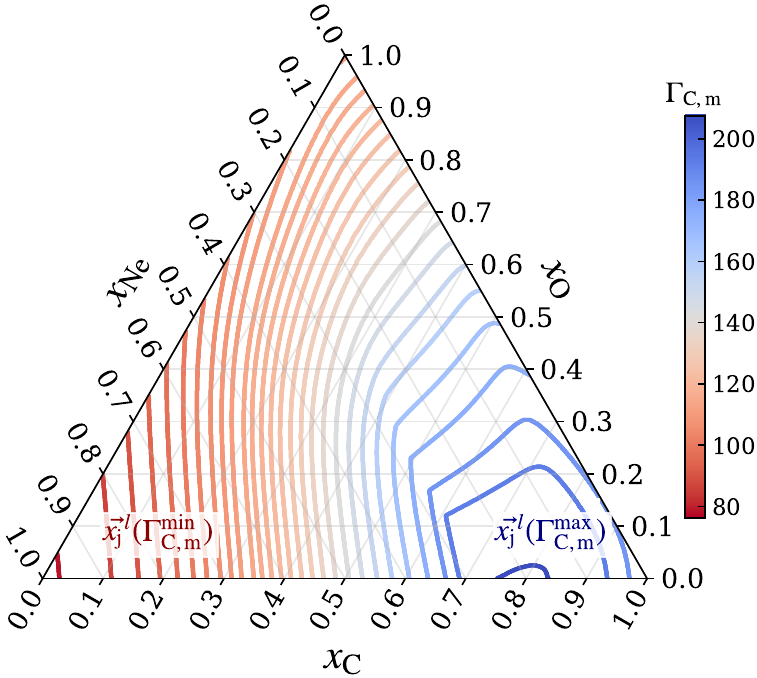}
  \caption{Phase diagram of three-component plasma for a C/O/Ne mixture constructed using the code of \citet{Caplan2018}. In the left and middle panels, we show, for a given coupling parameter $\Gamma_{\mathrm{C,m}}$,  the liquidus line $\vec{x}_{j}^{l}$ (green), the solidus line $\vec{x}_{j}^{s}$ (blue), and the unstable region, where the tie lines $\Delta{\vec{x}_{j}}$ (cyan) connect the solid and liquid points that are in equilibrium. The dashed line in the middle panel indicates a zone where a solid-solid coexistence occurs \citep[see][]{Caplan2020}, but we did not show their corresponding compositions since we assume monotonic cooling of the liquid, and only one zone at a time of homogeneous composition to transit from liquid to solid is needed for addressing WD crystallization. In the right panel, we show how the liquidus line varies for different values of $\Gamma_{\mathrm{C,m}}$.
  }
    \label{fig:CON_diagram}
\end{figure*}

\section{Phase separation for 3 species implementation}\label{sec:phase_sep}
In this section, we describe the implementation in the stellar evolution code MESA \citep {Paxton2011, Paxton2013, Paxton2015, Paxton2018, Paxton2019, Jermyn2023} of a 3-component phase diagram for studying the fractionation of elements during WD crystallization. 

\subsection{Three-component phase diagram}

Apart from C, O, and Ne, the other third most abundant elements predicted in the interior of UMWDs are Na and Mg \citep[e.g.][]{Camisassa2019, Camisassa2022a, Blouin2021c}. Thus, we calculated phase diagrams for the mixtures C/O/Ne, C/O/Mg, O/Ne/Na, and O/Ne/Mg to account for the whole range of predicted abundances. We used the code of \citet{Caplan2018}, which implements the semi-analytic method of \citet{MedinCumming2010} to construct the 3-component phase diagrams\footnote{\href{https://github.com/andrewcumming/phase\_diagram\_3CP}{https://github.com/andrewcumming/phase\_diagram\_3CP}}. This approach finds coexistence points in tangent planes of the minimum free energy surfaces, allowing coexisting solid-liquid abundances to be identified. The phase diagram depends on temperature, density, and charges $Z_{1}$, $Z_{2}$, and $Z_{3}$ in the particular combination given by the coupling parameters $\Gamma_{j,\mathrm{m}}=Z_{j}^{5/3}e^{2}/(a_{e}k_{B}T)$, where $e$ is the electron charge, $k_{B}$ the Boltzmann constant, and $a_{e}=3/(4\pi n_{e})$, with $n_{e}$ the electron number density. A one-component plasma transits from liquid to solid when this parameter exceeds a value of $\Gamma_{j,\mathrm{crit}}\approx178$ \citep{Potekhin2000, MedinCumming2010, Caplan2018, Bauer2020, Jermyn2021}. Thus, this approach only needs the ratio of how much this critical value is exceeded for a one-component plasma $\Gamma_{1,\mathrm{m}}/\Gamma_{1,\mathrm{crit}}$, and the charges of the components to construct the free-energy surfaces \citep[see][for more details]{MedinCumming2010, Caplan2018}.

\begin{figure*}
    \centering
    \includegraphics[width=0.5\textwidth]{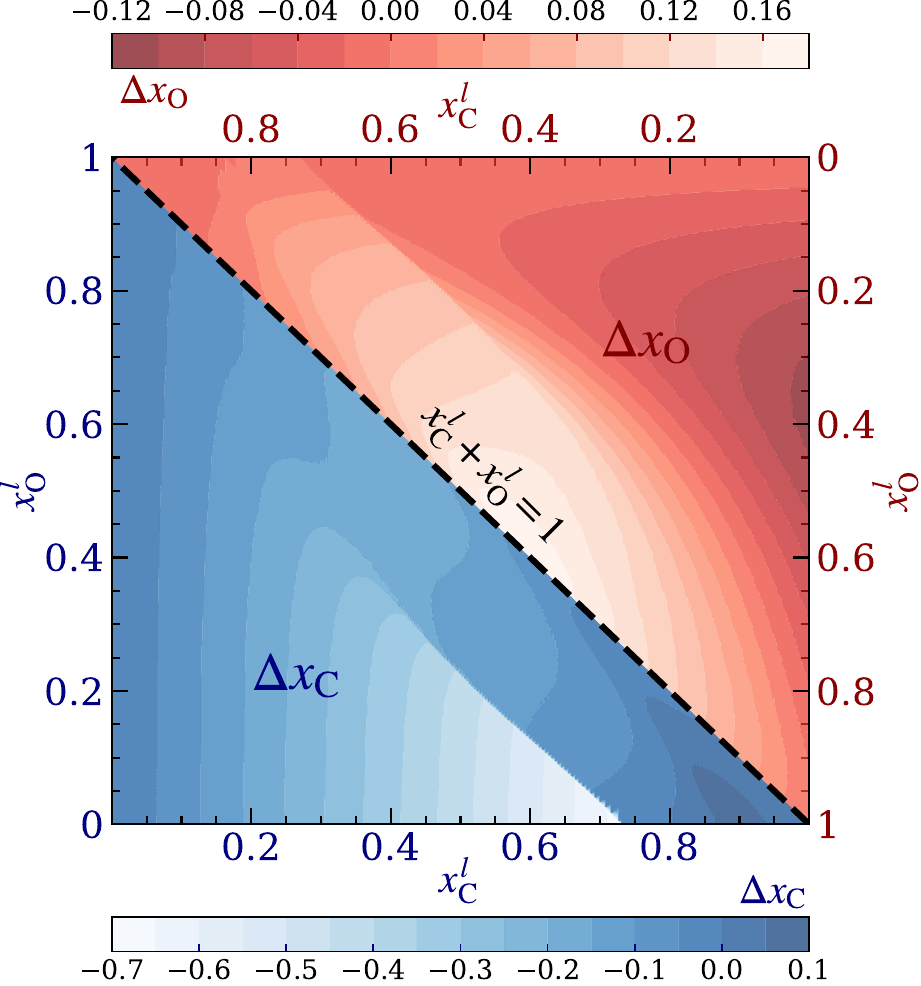}\hspace{0.1in}
    \includegraphics[width=0.48\textwidth]{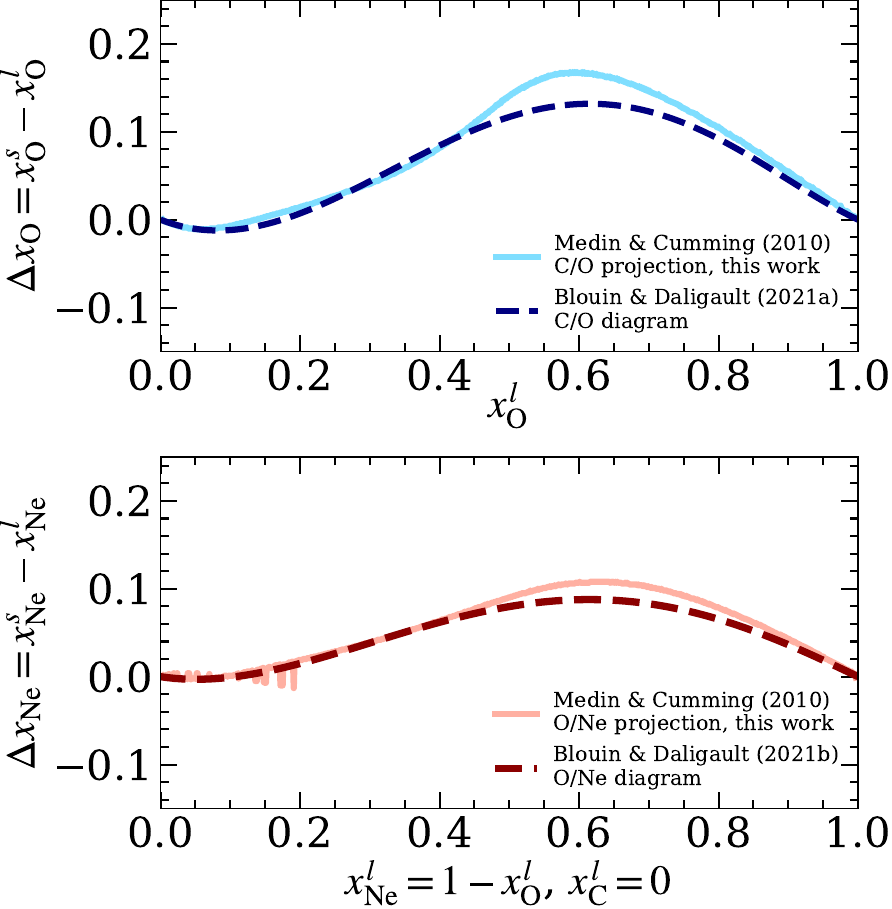}\hspace{0.in}
  \caption{The left panel shows color maps representing the concentration differences $\Delta{x_{\mathrm{C}}}$ and $\Delta{x_{\mathrm{O}}}$ as a function of $x^{l}_{\mathrm{C}}$ and $x^{l}_{\mathrm{O}}$  predicted from the C/O/Ne phase diagram. The upper right panel shows the two dimensional projection of $\Delta{x_{\mathrm{O}}}$ as function of $x^{l}_{\mathrm{O}}$ when $x^{l}_{\mathrm{Ne}}=0$ compared to the analytical formula for the C/O diagram of \citet{Blouin2021b}, while the lower right panel shows the projection of the O/Ne diagram with $\Delta{x_{\mathrm{Ne}}}$ as function of $x^{l}_{\mathrm{Ne}}$ considering  $x^{l}_{\mathrm{C}}=0$, compared to the analytical formula for the O/Ne diagram of \citet{Blouin2021c}.
  }
    \label{fig:Deltas}
\end{figure*}

In the first two panels of Figure \ref{fig:CON_diagram}, we show the phase diagram of the C/O/Ne mixture for two different values of $\Gamma_{\mathrm{C,m}}$. We show the liquidus line (green) composed of all the points $\vec{x}^{l}_{j}=(x^{l}_{\mathrm{C}},x^{l}_{\mathrm{O}},x^{l}_{\mathrm{Ne}})$ delimiting the pure liquid mixture, and the solidus (blue) for all the points $\vec{x}^{s}_{j}=(x^{s}_{\mathrm{C}},x^{s}_{\mathrm{O}},x^{s}_{\mathrm{Ne}})$ delimiting the solid phase, where $x_{j}$ is the number fraction of species $j$ (also referred as concentration). Visually, in Figure \ref{fig:CON_diagram}, all concentrations below the liquidus line remain liquid, and all above the solidus are pure solid. The tie lines connecting the solidus and liquidus (cyan lines) correspond to unstable regions where the mixture exists in a combination of both phases. However, the liquid-to-solid phase transition in WDs occurs in a thermally-stable background with small viscosity ($\mathrm{Pr} \sim 10^{-2}$, e.g. \citealt{NandkumarPethick1984, Isern2017, MontgomeryDunlap2024}), so the newly-formed crystals quickly sediment and the intermediate region does not last long in the star interior. Thus, during WD crystallization the regions that transit from liquid to solid suffer a change in composition given by the difference $\Delta\vec{x_{j}}=\vec{x}^{s}_{j}-\vec{x}^{l}_{j}$, while the remaining liquid changes according to $-\Delta\vec{x_{j}}$ to conserve number of ions.

In the phase diagrams of Figure \ref{fig:CON_diagram}, the parameter $\Gamma_{\mathrm{C,m}}$ is a proxy for the change in temperature as $\Gamma_{j,\mathrm{m}}\propto1/T$. In the third panel of Figure \ref{fig:CON_diagram} we show how the liquidus line changes for a wide range of $\Gamma_{\mathrm{C,m}}$, We note that as $\Gamma_{\mathrm{C,m}}$ increases the mixture can only be liquid for $\mathrm{C}$-rich abundances, and the liquidus line moves visually to the right corner ($x_{\mathrm{C}}=1$) of the diagram. Each composition can then be associated with a unique $\Gamma_{\mathrm{C,m}}$, which is the value of $\Gamma_{\mathrm{C,m}}$ at which the liquidus passes through that composition. Since  $x^{l}_{\mathrm{C}}+x^{l}_{\mathrm{O}}+x^{l}_{\mathrm{Ne}}=1$, $\Gamma_{\mathrm{C,m}}$ can be a function of only two concentrations. This is similar to what is presented by \citet{Blouin2021b, Blouin2021c} for the C/O and O/Ne phase diagrams. Following these references, we then construct the change in concentrations $\Delta{x_{\mathrm{C}}}$ and $\Delta{x_{\mathrm{O}}}$ as a function of $x^{l}_{\mathrm{C}}$ and $x^{l}_{\mathrm{O}}$. Note that the concentration change in Ne is just $\Delta{x_{\mathrm{Ne}}}=-(\Delta{x_{\mathrm{C}}}+\Delta{x_{\mathrm{O}}})$.

In the left panel of Figure \ref{fig:Deltas} we show the predicted $\Delta{x_{\mathrm{C}}}$ and $\Delta{x_{\mathrm{O}}}$ from the C/O/Ne phase diagram as a function of the concentrations $x^{l}_{\mathrm{C}}$ and $x^{l}_{\mathrm{O}}$, the difference in the concentrations are shown as color maps. Note that, since these numerical functions are limited by the domain $0\leq x^{l}_{\mathrm{C}}+x^{l}_{\mathrm{O}}\leq1$, the left and bottom axes (bottom left half of the plot) are used to show $\Delta{x_{\mathrm{C}}}$, while the right and top axes (upper right half of the plot) are used to show $\Delta{x_{\mathrm{O}}}$. Along the line $x^{l}_{\mathrm{C}}+x^{l}_{\mathrm{O}}=1$ the functions describe the C/O phase diagram; we take the values of the concentration differences at this line to construct $\Delta{x_{\mathrm{O}}}$ as a function of $x^{l}_{\mathrm{O}}$ to compare with the analytical fit of \citet{Blouin2021b} for $\Delta{x_{\mathrm{O}}(x^{l}_{\mathrm{O}})}$. We also take the results of $\Delta{x_{\mathrm{O}}}$ at $x^{l}_{\mathrm{C}}=0$ to construct $\Delta{x_{\mathrm{Ne}}(x^{l}_{\mathrm{Ne}})}$ and compare with the fit for the O/Ne phase diagram in \citet{Blouin2021c}. These analytical fits are compared to the results of our work on the right panel of Figure \ref{fig:Deltas}. We note that, in the limit of a two-component mixture, our numerical functions agree with the fits of \citet{Blouin2021b, Blouin2021c} that are currently used in the phase separation routine of MESA \citep{Bauer2023}. We proceed similarly to obtain numerical functions for the 3-component mixtures C/O/Mg, O/Ne/Na, and O/Ne/Mg.

\subsection{Implementation in the MESA phase separation routine}\label{sec:2.2}
To implement the 3-species phase transition in MESA, version r24.08.1, for species with concentrations $x_{1}$, $x_{2}$, and $x_{3}$, we first constructed tables mapping a grid of values of $(x^{l}_{\mathrm{1}}, x^{l}_{\mathrm{2}})$ to the corresponding $\Delta{x_{\mathrm{1}}}$ and $\Delta{x_{\mathrm{2}}}$ for all the mixtures considered\footnote{We directly modified the routine \texttt{phase\_separation.f90} in the main code and made available our extended version in a Zenodo repository \href{https://zenodo.org/records/15694294}{DOI:10.5281/zenodo.15694293}. We also provide the tables used for implementing the ternary phase diagrams and the inlists necessary to reproduce our UMWD models described in Section \ref{sec:WD_models}.}.

Next, following the work of \citet{Bauer2023}, at any given time, we identify the location of the mass coordinate for the phase transition in the current model. This is obtained using the phase parameter $\phi$ implemented in the Skye equation of state \citep{Jermyn2021}, which smoothly interpolates between the free energy of liquid and solid at each location $k$ of the star. This smoothed free energy allows for the computation of the latent heat from crystallization in the energy equation. Formally, the transition from liquid to solid would be at $\phi=0.5$, since it is the exact point where the free energy of both phases coincides. However, the latent heat calculation in Skye does not account for the composition changes from solidification. To avoid energy miscalculations in latent heat from composition changes, \citet{Bauer2023} implements the fractionation of the elements separately at $\phi=0.9$, and we adopt the same approach in this work.

In MESA, phase separation is computed as follows: the location of the crystal core is set to be initially $m_{\mathrm{cr}}=0$, the cells $k$ which fulfill $\phi_{k}\geq0.9$ phase separate if $m_{\mathrm{cr}}<m_{k}$ in the current time step. The cells in the code are considered increasing from the surface to the center of the star, then $k=1$ is at the surface, and $k=N_{z}$ is the core. The iteration of the phase value continues outwards. If $\phi_{k}<0.9$, then the new crystal core is defined at $m_{\mathrm{cr}}=m_{k+1}$ for the following time step. The cells that transited from liquid to solid redistribute the elements into the liquid region above according to the $\Delta{x_{j}}$ functions.

Since in MESA, the chemical abundances are in mass fractions, the composition changes for the 3-component implementation are obtained by converting $X_{1,k}$ and $X_{2,k}$ to number fractions $x_{1,k}$ and $x_{2,k}$, then we make interpolations in our tables to obtain the concentration differences $\Delta{x_{1,k}}(x_{1,k},x_{2,k})$ and $\Delta{x_{2,k}(x_{1,k},x_{2,k})}$. The new number concentrations are $x_{1,\mathrm{new}}=x_{1,k}+\Delta{x_{1,k}}$ and $x_{2,\mathrm{new}}=x_{2,k}+\Delta{x_{2,k}}$, which are converted again to mass fractions $X_{1,\mathrm{new}}$ and $X_{2,\mathrm{new}}$ to finally obtain $\Delta{X_{1,k}}=X_{1,\mathrm{new}}-X_{1,k}$ and $\Delta{X_{2,k}}=X_{2,\mathrm{new}}-X_{2,k}$. Thus, the corresponding compositions updated in the cell  $m_{k}$ that phase separates are similar to equations (3) and (4) of \citet{Bauer2023},
\begin{equation}
    X_{1,k}\rightarrow{X_{1,k}+\Delta{X_{1,k}}},
\end{equation}
\begin{equation}
    X_{2,k}\rightarrow{X_{2,k}+\Delta{X_{2,k}}},
\end{equation}
and for the third element, we get
\begin{equation}
    X_{3,k}\rightarrow{X_{3,k}-(\Delta{X_{1,k}}+\Delta{X_{2,k}})}.
\end{equation}

The composition is also updated in the zone above the new solid boundary in such a way as to conserve the mass of each species, so the cell above ($k-1$) changes composition according to
\begin{equation}
    X_{1,k-1}\rightarrow{X_{1,k-1}-\Delta{X_{1,k}}\delta{m_{k}}/\delta{m_{k-1}}},
\end{equation}
\begin{equation}
    X_{2,k-1}\rightarrow{X_{2,k-1}-\Delta{X_{2,k}}\delta{m_{k}}/\delta{m_{k-1}}},
\end{equation}
and
\begin{equation}
    X_{3,k-1}\rightarrow{X_{3,k-1}+(\Delta{X_{1,k}}+\Delta{X_{2,k}})\delta{m_{k}}/\delta{m_{k-1}}},
\end{equation}
where we have accounted for difference between the mass $\delta{m_{k}}$ contained in the cell $k$ relative to the mass $\delta{m_{k-1}}$ in the contiguous cell $k-1$ \citep{Bauer2023}. Note that, since our tables are constructed to satisfy $x_{1}+x_{2}+x_{3}=1$, the input mass fractions $X_{1}$ and $X_{2}$ must be normalized by $\sum^{3}_{j=1}X_{j}$ before interpolating, and then rescaled when the compositions in cells $k$ and $k-1$ are modified.

At any given time, we use the 3-species phase diagram corresponding to the three most abundant elements at the solidification front. Thus, we do not have to specify which table to use a priori, and we only have to indicate the option \texttt{phase\_separation\_option =`3c'} in the inlists. We only considered phase separation when the condition $X_{1}+X_{2}+X_{3}>0.7$ is fulfilled. This is to turn off the phase separation at the base of the atmosphere of the WD.

\begin{figure*}
    \centering
    \includegraphics[width=0.99\textwidth]{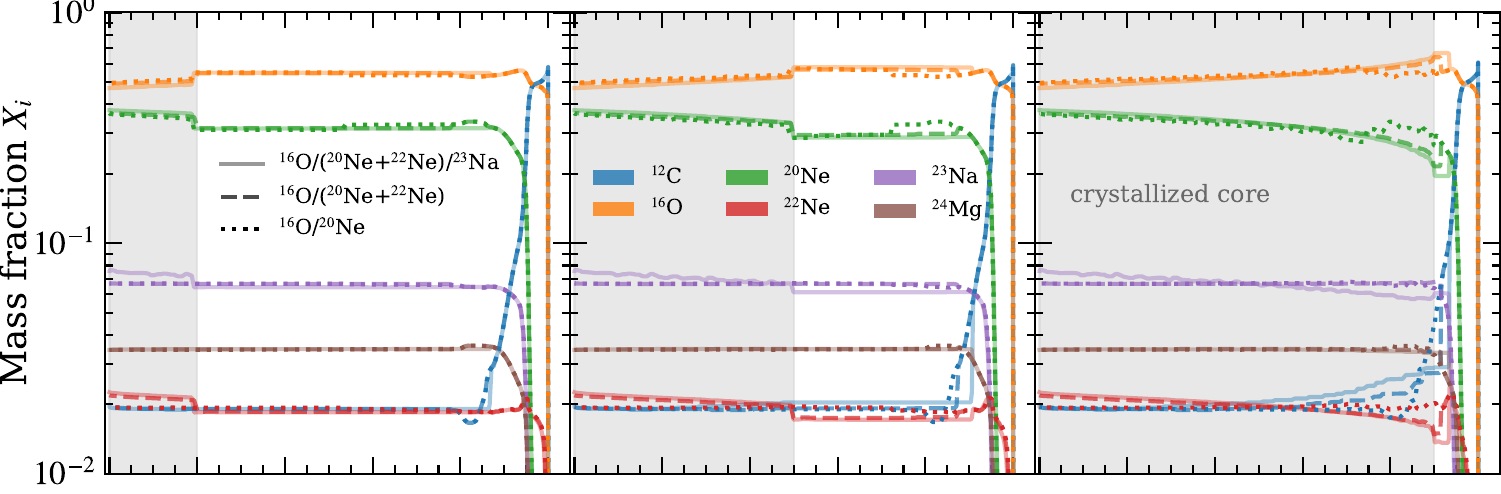}
    \includegraphics[width=0.99\textwidth]{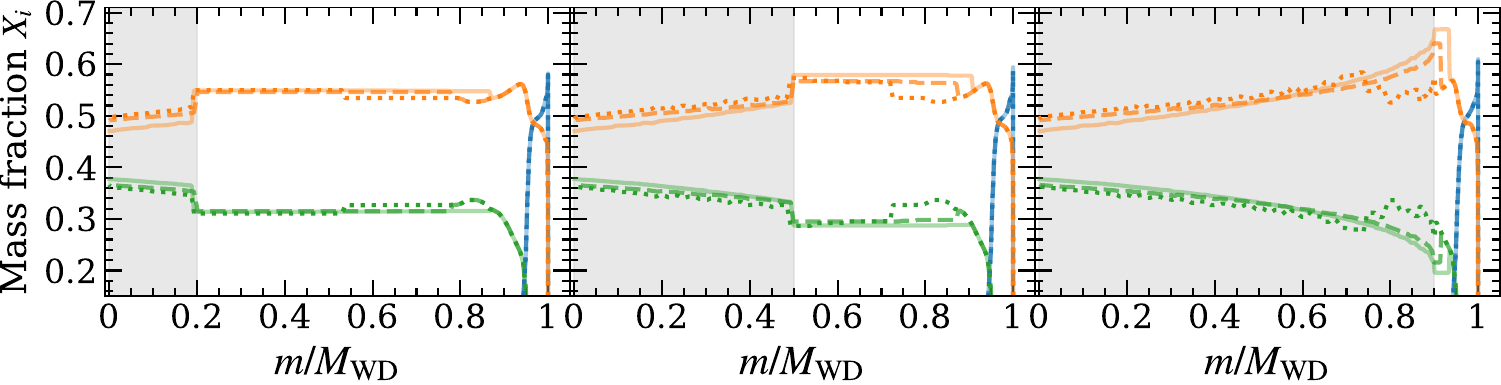}
  \caption{Composition profiles, mass fraction abundances $X_{i}$, as a function of the mass coordinate $m/M_{\mathrm{WD}}$, for different species at three different stages of crystallization for $1.1\ M_{\odot}$ O/Ne/Na WD models. The solid lines show the chemical evolution when applying our 3-species phase separation, the dashed lines show the case of 2-species, including the $\mathrm{^{22}Ne}$ isotope, and the dotted lines show the simplest 2-species with only $\mathrm{^{16}O}$ /$\mathrm{^{20}Ne}$. The bottom panels zoom into the $\mathrm{^{16}O}$ and $\mathrm{^{20}Ne}$ compositions for the different cases in the top panels.
  }
    \label{fig:ONeNa_sep}
\end{figure*}

\begin{figure*}
    \centering
    \includegraphics[width=0.99\textwidth]{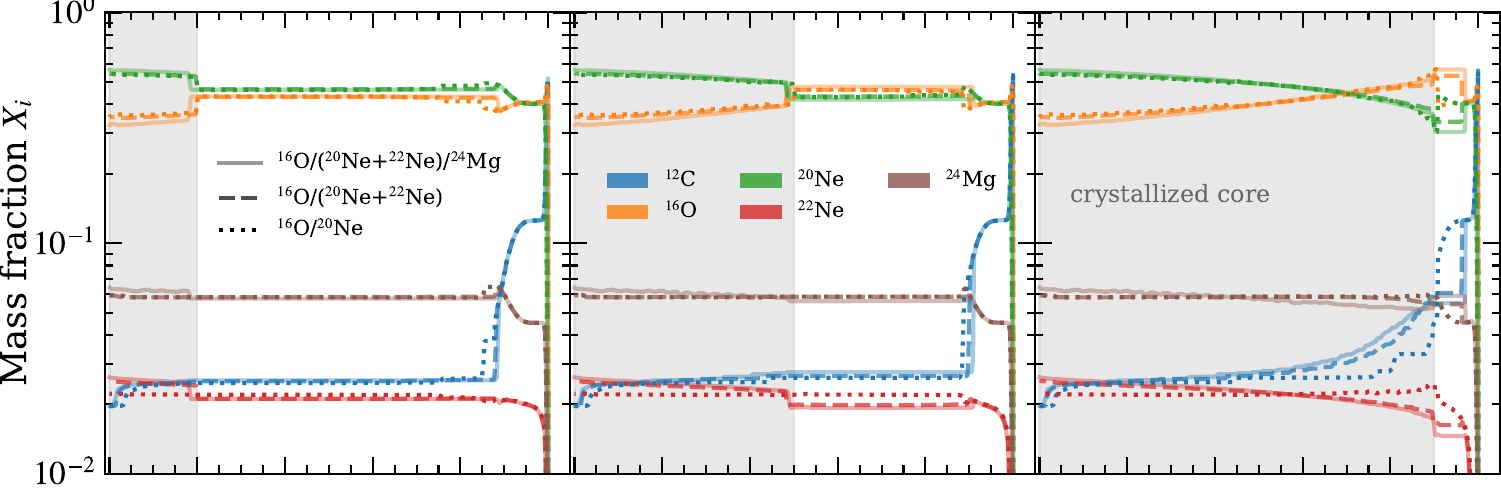}

    \includegraphics[width=0.99\textwidth]{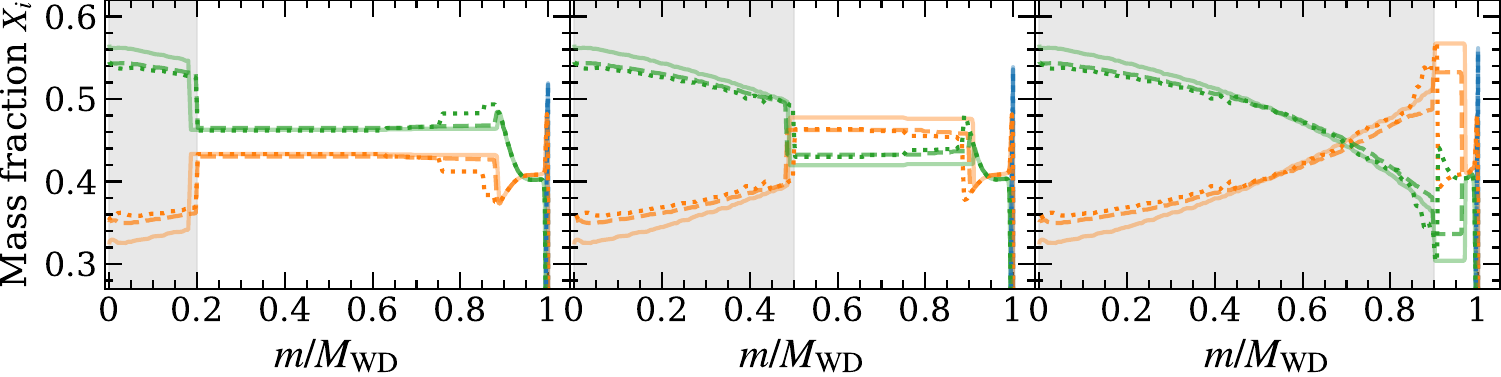}
  \caption{Composition profiles, mass fraction abundances $X_{i}$, as a function of the mass coordinate $m/M_{\mathrm{WD}}$, for different species at three different stages of crystallization for $1.1\ M_{\odot}$ Ne/O/Mg WD models. The solid lines show the chemical evolution when applying our 3-species phase separation, the dashed lines show the case of 2-species, including the $\mathrm{^{22}Ne}$ isotope, and the dotted lines show the simplest 2-species with only $\mathrm{^{16}O}$ /$\mathrm{^{20}Ne}$. The bottom panels zoom into the $\mathrm{^{16}O}$ and $\mathrm{^{20}Ne}$ compositions for the different cases in the top panels.
  }
    \label{fig:ONeMg_sep}
\end{figure*}

Note that while the calculation of the phase diagrams depends on the charge $Z_{j}$ of the species, it does not depend on the atomic number $A_{i}$. Therefore, when Ne ($Z=10$) is considered in the phase separation, we take as input the isotopes $X_{\mathrm{^{20}Ne}}$ and $X_{\mathrm{^{22}Ne}}$ separately to be converted to number fraction, and then $x_{\mathrm{Ne}}=x_{\mathrm{^{20}Ne}}+x_{\mathrm{^{22}Ne}}$ enters into the interpolation as a single specie. When converting back to mass fraction we rescale $\Delta{x_{\mathrm{Ne}}}(x_{\mathrm{^{iso}Ne}}/x_{\mathrm{Ne}})$ for $\mathrm{iso}=20,22$ to calculate the new values $x_{\mathrm{^{20}Ne,new}}$, $x_{\mathrm{^{22}Ne,new}}$, $X_{\mathrm{^{20}Ne,new}}$, and $X_{\mathrm{^{22}Ne,new}}$. Thus, the Ne abundance changes in cells $k$ and $k-1$ are obtained as follows

\begin{equation}
    X_{\mathrm{Ne},k}\rightarrow{\begin{cases} 
     X_{\mathrm{^{20}Ne},k}+\Delta{X_{\mathrm{^{20}Ne},k}}\\
    X_{\mathrm{^{22}Ne},k}+\Delta{X_{\mathrm{^{22}Ne},k}}
   \end{cases}}
\end{equation}

\begin{equation}
    X_{\mathrm{Ne},k-1}\rightarrow{\begin{cases} 
     X_{\mathrm{^{20}Ne},k-1}-\Delta{X_{\mathrm{^{20}Ne},k}}\delta{m_{k}}/\delta{m_{k-1}}\\
    X_{\mathrm{^{22}Ne},k-1}-\Delta{X_{\mathrm{^{22}Ne},k}}\delta{m_{k}}/\delta{m_{k-1}}.
   \end{cases}}.
\end{equation}
Note that if $\Delta{X_{3,k}}=\Delta{X_{\mathrm{Ne,k}}}=\Delta{X_{\mathrm{^{22}Ne},k}}+\Delta{X_{\mathrm{^{20}Ne},k}}$ we still have
$\Delta{X_{3,k}}=-(\Delta{X_{1,k}}+\Delta{X_{2,k}})$. We also included the fractionation of the isotope $^{22}\mathrm{Ne}$ into the two-species phase separation in MESA as the current version only accounts for $^{20}\mathrm{Ne}$ in the O/Ne crystallization.

The fluid mixing in the liquid phase and phase separation heating remain unmodified, and the details of these processes are described in sections 3.2 and 3.3 of \citet{Bauer2023}. We applied a minor fix to the evaluation of the mass of the crystallized core $m_{\mathrm{cr}}$ when the phase value in the $k=N_{z}$ (core) cell fulfills $\phi_{N_{z}}<0.9$. In the current implementation if  $m_{\mathrm{cr}}>0$, but $\phi_{N_{z}}<0.9$, the mass of the crystallization core is set again to $m_{\mathrm{cr}}=0$, which is not physically correct, as it implies that the core is being melted. This is due to the heating term from the phase separation, which is added in the next timestep as the difference in the internal energy upon the redistribution of the components. The difference in the internal energy pushes $\phi_{N_{z}}$ to a value smaller than 0.9 for a short time, and then it rises to $>0.9$ as the core cools, introducing an artificial extra fractionation as the only transition considered is liquid to solid in the code. To prevent this, we no longer decrease $m_{\mathrm{cr}}$ when  $\phi_{N_{z}}$ falls below 0.9 if the value of $m_{\mathrm{cr}}$ is already non-zero. This causes the mass of the crystal to remain unchanged until $\phi_{N_{z}}$ once again increases to above 0.9. We discuss this fix in more detail in Appendix \ref{appA}.

\section{Application to White Dwarf Models}\label{sec:WD_models}
\subsection{UMWD models}
We constructed a set of UMWDs following the methods of \citet{Lauffer2018}. Specifically, we evolved stars from the pre-main sequense to the WD stage increasing the amount of mass loss in the red giant branch (RGB) and AGB, and modifying the overshoot for the lower boundary of the metal burning convective region. The initial masses of our models were between $M_{\mathrm{ini}}=8\ M_{\odot}$ and $9.6\ M_{\odot}$, which produced WDs with masses from $M_{\mathrm{WD}}=1.06\ M_{\odot}$ to $1.35\ M_{\odot}$. We carried out this process for two different nuclear networks: first, the one used in \citet{Lauffer2018}, \texttt{co\_burn\_plus.net}, and second, the one used in \citet{Bauer2020}, \texttt{sagb\_NeNa\_MgAl.net} which includes the isotope $^{23}\mathrm{Na}$ as one of the products of carbon burning along with $^{20}$Ne and $^{24}$Mg. This difference in the nuclear network produces significant differences in the internal chemical profile. The first case produces models that have $^{24}$Mg as the third most abundant element for $M_{\mathrm{WD}}\gtrsim1.1\ M_{\odot}$, while for the other case $^{23}\mathrm{Na}$ is the third most abundant.

Although in both cases $^{16}$O and $^{20}$Ne are the two most abundant species in the core, the \citet{Lauffer2018} method mostly produces a core where $^{16}\mathrm{O}<^{20}$Ne, while the more detailed network generates a core where $^{16}\mathrm{O}>^{20}\mathrm{Ne}$ as in the case of \citet{Siess2006} and \citet{Camisassa2019}. Hereafter, we will refer to the set of models with $^{23}\mathrm{Na}$ as O/Ne/Na WDs, while the ones with $^{24}\mathrm{Mg}$ as the third most abundant species as the Ne/O/Mg WDs.  Additionally, given the enhanced mass loss during the AGB stage, the He and hydrogen (H) in the envelope are completely lost when the WD formed, and so we added a He atmosphere by accreting $10^{-4}\ M_{\odot}$ of He at the beginning of the cooling \citep[similar to the method of][]{Bauer2020}. This is done as current accurate models of the AGB show that He and/or H layers remain by the end of such phase \citep{Siess2007, Siess2010}.

\begin{figure}
    \centering
    \includegraphics[width=0.48\textwidth]{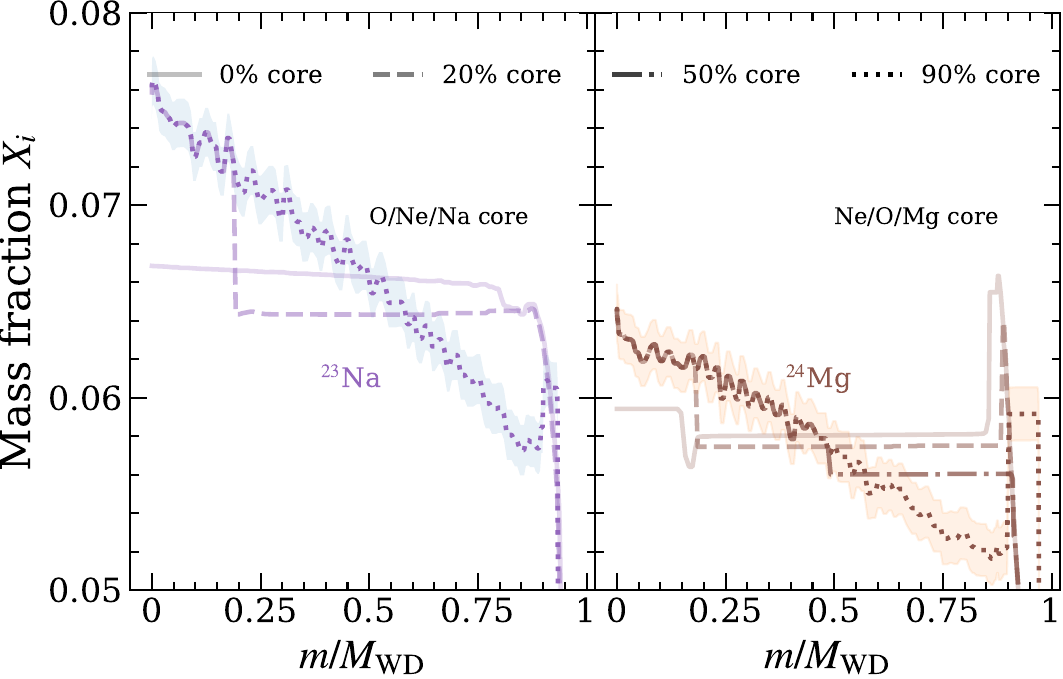}
  \caption{Mass fraction abundances $X_{i}$ of $\mathrm{^{23}Na}$ and $\mathrm{^{24}Mg}$, as a function of the mass coordinate $m/M_{\mathrm{WD}}$, at different stages of crystallization for a $1.1\ M_{\odot}$ O/Ne/Na (left panel) and Ne/O/Mg (right panel) WD model, respectively, when applying the 3-species phase separation. The error bars are computed considering the $X_{i}$ obtained from the range of concentrations $x_{i}\pm0.001$, with $0.001$ the concentration resolution in our phase diagrams.
  }
    \label{fig:Na_Mg}
\end{figure}

We applied the three-species and two-species phase separation to the O/Ne/Na and the Ne/O/Mg WD models. In Figure \ref{fig:ONeNa_sep} and \ref{fig:ONeMg_sep}, we show the evolution of the chemical profiles for the case of a crystallizing O/Ne/Na and Ne/O/Mg $1.1\ M_{\odot}$ WD, respectively. The solid lines show the case when applying the three-component phase separation, the dashed lines show the two-component phase separation including the isotope $^{22}\mathrm{Ne}$, and the dotted lines show the two-component separation for $^{16}\mathrm{O}/^{20}\mathrm{Ne}$. Note that, as the crystallization advances, in the two-component phase separation case with $^{22}\mathrm{Ne}$, the abundance profiles of the species $^{16}\mathrm{O}$ and $^{20}\mathrm{Ne}$ become more similar to the three-component separation than the case without $^{22}\mathrm{Ne}$. 

Although the same phase diagram fitting formula from \citet{Blouin2021c} is used in the two-component separation with and without $^{22}\mathrm{Ne}$, it seems that the inclusion of this isotope in the phase separation generates a significant fluid mixing in the liquid phase, even when its local mass abundance across the WD is of an order of $10^{-2}$. This enhanced mixing can be observed in Figures \ref{fig:ONeNa_sep} and \ref{fig:ONeMg_sep}  as the more extended flat profiles for $^{16}\mathrm{O}$ and $^{20}\mathrm{Ne}$ in the liquid phase, for the phase separation cases that include $^{22}\mathrm{Ne}$. This occurs because the phase diagram only depends on the charge $Z_{j}$ of the species, and then, when separating the $\mathrm{Ne}$, the relative change in composition of $^{22}\mathrm{Ne}$ will be higher in terms of mass fraction than $^{20}\mathrm{Ne}$, due to the extra neutrons in $^{22}\mathrm{Ne}$, while keeping the same concentration change of both isotopes due to the phase diagram shape. 

When computing the phase diagram, we use a resolution of $0.001$ in the concentrations, which introduces discretization errors of that size when computing the $\Delta{x_{j}}$. In Figure \ref{fig:Na_Mg}, we zoom in the evolution of $^{23}\mathrm{Na}$ and $^{24}\mathrm{Mg}$ abundances for the O/Ne/Na and Ne/O/Mg models. The short timescale variations that can be seen in the curves for $^{23}\mathrm{Na}$ and $^{24}\mathrm{Mg}$ abundances are a numerical artifact coming from the discretization of composition $x^{l}_{j}$ when searching for tangent points in the diagrams. Making an analytical surface fit for the numerical functions we constructed with the tables could help smooth this behavior on the abundance profile. We added in the figure a shaded area equivalent to an abundance number fraction $x_{j}\pm{0.001}$ for the profiles with $90\%$ of the crystallized core. This shows that,
despite the errors, the initial composition profiles change significantly as crystallization progresses. Therefore, the errors do not notably affect the overall trend of compositional change upon fractionation. The evolution of the compositions is such that the solid phase is enriched in $^{23}\mathrm{Na}$ and $^{24}\mathrm{Mg}$, respectively, while depleted in the liquid. The value $0.001$ was selected as the error in the measurement because it is the spacing in our tables for the concentrations in the liquidus.
\subsection{Crystallization-driven convection for 3-component phase separation}

Crystallizing WDs may experience convection in their interior even when they have a thermally stable/subadiabatic background. The fractionation of the components upon the phase transition generates a compositionally unstable region \citep[e.g.][]{Isern2017, Fuentes2023}. Under this premise, in this section, we show how using mixing length theory (MLT) and implementing the 3-component phase separation can modify the convective efficiency during the crystallization of WDs. 

As pointed out in previous works \citep{MedinCumming2015, Fuentes2023, Castro-Tapia2024}, in MLT, the buoyant acceleration and the convective velocity are proportional to the density contrast $D\rho$ between a fluid element and its surroundings after it has moved through a mixing length $\ell$. This quantity can be written in terms of the temperature and composition gradients for an $n$-component mixture as follows
\begin{align}\label{eq_del_rho_comp}
    \nonumber\frac{D\rho}{\rho}&=-\frac{\chi_{T}}{\chi_{\rho}}\frac{DT}{T}-\frac{1}{\chi_{\rho}}\sum_{i=1}^{n-1}\chi_{{i}}\frac{DX_{i}}{X_{i}}~,\\
    &=-\frac{\chi_{T}}{\chi_{\rho}}(\nabla-\nabla_{e}+\nabla_{\mathrm{com}})\frac{\ell}{2H_{P}}~,
\end{align}
where $\chi_T = \left.\partial\ln P/\partial\ln T\right|_{\rho, X}$,  $\chi_\rho = \left.\partial\ln P/\partial\ln \rho\right|_{T, X}$, $DT=T(\nabla-\nabla_{e})(\ell/2H_{P})$ is the temperature contrast, with $H_{P}$ the pressure scale height, $\nabla=d\ln{T}/d\ln{P}|_{\star}$ the temperature gradient with pressure in the star, and $\nabla_{e}$ the temperature gradient with pressure experienced by a fluid element as it moves. The subscript $X$ indicates that the composition is kept constant for the partial derivative.

The composition contrast of each component with mass fraction $X_{i}$ is written as $DX_{i}=X_{i}\nabla_{X_{i}}(\ell/2H_{P})$, where $\nabla_{X_{i}}=d\ln{X_{i}}/d\ln{P}|_{\star}$. However, to write the total composition gradient given the summation over the $n$ species, we defined $\nabla_{\mathrm{com}}=\chi_{T}^{-1}\sum_{i=1}^{n-1}\chi_{i}\nabla_{X_{i}}$, where we have written $\chi_{i}$ to differentiate between $\chi_{X_{i}}=\left.\partial\ln P/\partial\ln X_{i}\right|_{\rho, T, X_{j\neq i},Y_{e}}$ with $j=1,2...,i-1,i+1,...,n$, and $\chi_{i}=\left.\partial\ln P/\partial\ln X_{i}\right|_{\rho, T, X_{j\neq i}}$ with $j=1,2...,i-1,i+1,...,n-1$, which comes from imposing the mass conservation $\sum_{i=1}^{n}X_{i}=1$. Thus,  defining the electron fraction $Y_{e}=\sum_{i=1}^{n}Y_{i}X_{i}$, where $Y_{i}=Z_{i}/A_{i}$ is the electron fraction of species $i$ with atomic number $Z_{i}$ and mass number $A_{i}$, we proceed as in the Appendix A of \citet{MedinCumming2015} and define

\begin{equation}\label{chi_i}
    \chi_{i}=\chi_{X_{i}}-\chi_{X_{n}}\frac{X_{i}}{X_{n}}+\chi_{Y_{e}}\frac{(Y_{i}-Y_{n})X_{i}}{Y_{e}},
\end{equation}
where $\chi_{Y_{e}}=\left.\partial\ln P/\partial\ln Y_{e}\right|_{\rho, T, X}$.

In \citet{Castro-Tapia2024}, we used the composition and heat fluxes into the fluid region rather than the gradients to characterize the convective instability given by $D\rho/\rho<0$. The composition flux $F_{X_{i}}$ of each species from its composition contrast is given by
\begin{equation}
    F_{X_{i}}=\rho v_{c}DX_{i}=\rho v_{c}X_{i}\nabla_{X_{i}}\frac{\ell}{2H_{P}},
\end{equation}
where $v_{c}$ is the convective velocity.
Dividing by $X_{i}$ and multiplying by $(\chi_{i}/\chi_{T})$ we get $\chi_{T}^{-1}(\chi_{i}F_{X_{i}}/X_{i})= \rho v_{c}\chi_{T}^{-1}\chi_{i}\nabla_{X_{i}}(\ell/2H_{P}).$
Summing over the $n$ species and taking the definition of the convective efficiency $\Gamma=(1/2a_{0})(v_{c}\ell/\kappa_{T})$, where $a_{0}$ is a geometric term that depends on the assumptions for the shape of the fluid element, and $\kappa_{T}=4acT^{3}/(3\kappa \rho^{2}c_{P})$ is the thermal diffusivity, where $\kappa$ is the opacity, $a$ the radiation constant, and $c$ the speed of light, we get a relation between the composition fluxes and the total composition gradient
\begin{equation}
    \chi_{T}^{-1}\sum_{i=1}^{n-1}\frac{\chi_{i}F_{X_{i}}}{X_{i}}= \frac{\rho \kappa_{T}}{H_{P}} a_{0}\Gamma\nabla_{\mathrm{com}}.
\end{equation}
This relation allows us to define the following dimensionless parameter
\begin{equation}
     \tau=\chi_{T}^{-1}\sum_{i=1}^{n-1}\frac{\chi_{i}F_{X_{i}}}{X_{i}}\frac{H_{p}}{\rho\kappa_{T}\nabla_{\mathrm{ad}}}= a_{0}\Gamma\frac{\nabla_{\mathrm{com}}}{\nabla_{\mathrm{ad}}},
\end{equation}
which is equivalent to equation (21) of \citet{Castro-Tapia2024}, but generalized for $n$ species instead of just 2 as used in that work.

We can relate the thermal gradients to the total heat flux as follows
\begin{equation}\label{eq_Ftot}
 F=F_{\mathrm{rad}}+F_c=\rho c_{P} \kappa_{T}\frac{T}{H_{P}}\nabla_{\mathrm{rad}}~,
\end{equation}
where $\nabla_{\mathrm{rad}}$ is the temperature
gradient needed to transport all the energy by radiative diffusion/conduction, $F_{\mathrm{rad}}=\rho c_{P} \kappa_{T}(T/H_{P})\nabla$ is the diffusive heat flux, and $ F_c = \rho v_c c_{P} DT = \rho v_c c_{P} T\left(\nabla-\nabla_e\right)(\ell/2H_{P})$ is the heat flux carried by convection. Now, taking the definition of convective efficiency in terms of the thermal losses as the fluid element moves due to convection, $ \Gamma =  (\nabla-\nabla_{e})/(\nabla_{e}-\nabla_{\mathrm{ad}}) = (1/2a_{0})(v_{c}\ell/\kappa_{T})$,
we can get the same MLT described in \citet{Castro-Tapia2024}, and the density contrast in terms of the fluxes is given by
\begin{equation}\label{eq_del_rho_comp}
   \frac{D\rho}{\rho}=-\frac{\chi_{T}}{\chi_{\rho}}\frac{\nabla_{\mathrm{ad}}}{a_{0}\Gamma}\left[\left(\frac{\nabla_{\mathrm{rad}}}{\nabla_{\mathrm{ad}}}-1\right)\zeta+\tau\right]\frac{\ell}{2H_{P}}~,
\end{equation}
where $\zeta=a_{0}\Gamma^{2}/[1+\Gamma(1+a_{0}\Gamma)]$.

\begin{figure*}
    \includegraphics[width=0.999\textwidth]{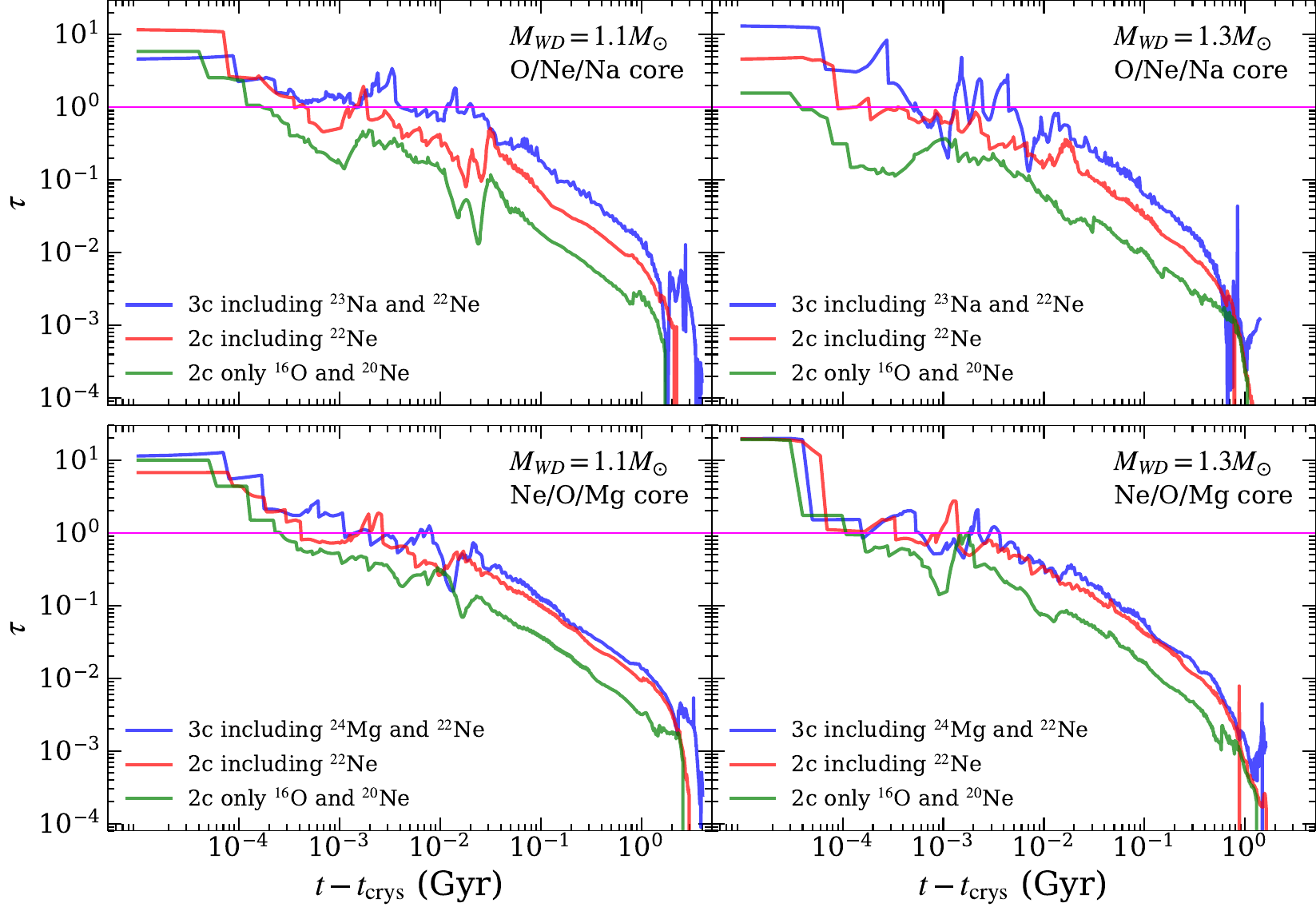}
  \caption{Buoyancy flux $\tau$ induced by a composition flux in the liquid phase as crystallization advances, derived from the O/Ne UMWDs models used in this work. The top panels show models for O/Ne/Na cores, while the bottom panels show the Ne/O/Mg cores. The magenta line shows the location of $\tau=1$, where convection efficiency transits from fast ($\tau>1$) to slow (0<$\tau<1$). 3c stands for 3-component, while 2c stands for 2-component to differentiate the phase separation treatment. 
  }
    \label{fig:tau_3c}
\end{figure*}

This generalizes the previous treatment of \citet{Castro-Tapia2024} to multiple species.  Whereas previously $\tau$ was referred to as measuring the composition flux, we see here that in fact $\tau$ can be understood as buoyancy flux since it characterizes the density contrast and it is directly proportional to the total composition gradient $\nabla_{\mathrm{com}}$. The same occurs with the term $\nabla_{\mathrm{rad}}/\nabla_{\mathrm{ad}}-1$ which is related to the thermal contrast with the gradient $\nabla-\nabla_{e}$. Thus, as pointed out by \citet{Castro-Tapia2024}, since the buoyant acceleration is proportional to $-g\ell(D\rho/\rho)$, with $g$ the local gravity, $\tau>0$ and $\nabla_{\mathrm{rad}}/\nabla_{\mathrm{ad}}>1$ contribute to the development of an unstable region. Additionally, when considering the scaling $v_{c}^{2}\approx-g(D\rho/\rho)(\ell/4)$, they found that the convective efficiency can be solved for given values of the fluxes, which for a subadiabatic regime $\nabla_{\mathrm{rad}}/\nabla_{\mathrm{ad}}<1$, is modulated by the value of $\tau$. Specifically, $0<\tau<1$ gives $\Gamma\ll1$, which is in the thermohaline convection regime, while $\tau>1$ gives an overturning convection regime with $\Gamma\gg1$. These regimes remain even when the scaling for $v_{c}$ considers rotation and/or magnetic fields \citep{Fuentes2023, Fuentes2024, Castro-Tapia2024b}.

In Figure \ref{fig:tau_3c} we show the predicted value of $\tau$ for different cases of the phase separation implementation in UMWDs. To estimate the individual composition fluxes $F_{X_{i}}$, we followed the temporal growth of the crystal core and computed
\begin{equation}
    F_{X_{i}}=\frac{\dot{M}_{\mathrm{c}}\Delta{X_{i}}}{4\pi R_{\mathrm{c}}^{2}},
\end{equation}
where $M_{\mathrm{c}}$ and $R_{\mathrm{c}}$ are the mass and radius of the solid core obtained as the outermost location that fulfills $\phi\geq0.9$ at a given time; $\Delta{X_{i}}$ is the mass abundance difference of the component $i$ that goes into the liquid as fractionation occurs. This difference is obtained using the interpolation in our tables for the 3-species phase diagrams as explained in Section \ref{sec:2.2} and the fitting formula of \citet{Blouin2021c} for the concentrations when applying the 2-species cases. The upper panels of Figure \ref{fig:tau_3c} show the models for $1.1M_{\odot}$ and $1.3M_{\odot}$ O/Ne/Na WDs, while the lower panels show models for the same masses, but with Ne/O/Mg cores. We can note that the inclusion of 3 species in the phase separation does modify the value of the buoyancy flux, making not only $\tau$ larger than in the two-species case for most of the cooling time, but also prolonging the time for which $\tau>1$. This feature is more significant in the case of the O/Ne/Na core. 

An increase in the value of $\tau$ also appears when the isotope $^{22}\mathrm{Ne}$ is in the 2-species phase separation. Moreover, this inclusion by itself can increase $\tau$ to values comparable to the 3-species phase separation for the Ne/O/Mg cores. In general, neutron-rich species contribute larger buoyant fluxes. This occurs because the phase diagrams depend on the Coulomb interactions, which do not depend on the mass numbers $A_{i}$. When separating an equal concentration of an element (for example, Ne) into the solid or liquid, the more massive isotopes will increase the density contrast more. We can understand this from the last term in Equation \eqref{chi_i}, where for a neutron-rich abundance $X_{i}$, the value $(Y_{i}-Y_{n})\neq0$, and the corresponding $\chi_{i}$ will increase in magnitude. Consequently, neutron-rich species also have a larger contribution to the composition gradient. This would also explain the more extended zone with flat composition in the liquid phase when comparing the cases in Figures \ref{fig:ONeNa_sep} and \ref{fig:ONeMg_sep}. The mixing in the liquid zone for phase separation is handled in MESA by measuring the local composition gradient and using the Ledoux criterion to establish to which zone the composition must be flattened/mixed. A larger composition gradient means that more mixing in the liquid is necessary to stabilize the Ledoux term.

The value of a buoyancy flux $\tau>1$ for a longer time when adding more species to phase separation can potentially help to power a dynamo due to efficient convection \citep{Fuentes2024} in UMWDs. As pointed out by \citet{Castro-Tapia2024b}, the evolution of such a magnetic field will depend on the size of the convection zone during $\tau>1$. They found that a larger zone enables more rapid transport of the field to the surface, avoiding freezing of the field in the crystal core. The more extended mixing zone observed in Figures \ref{fig:ONeNa_sep} and \ref{fig:ONeMg_sep} for 3-species  phase separation may also favor a crystallization-driven dynamo to explain some magnetic UMWDs.

\subsection{Implications for the Cooling Curve}
\begin{figure*}
    \centering
    \includegraphics[width=0.99\textwidth]{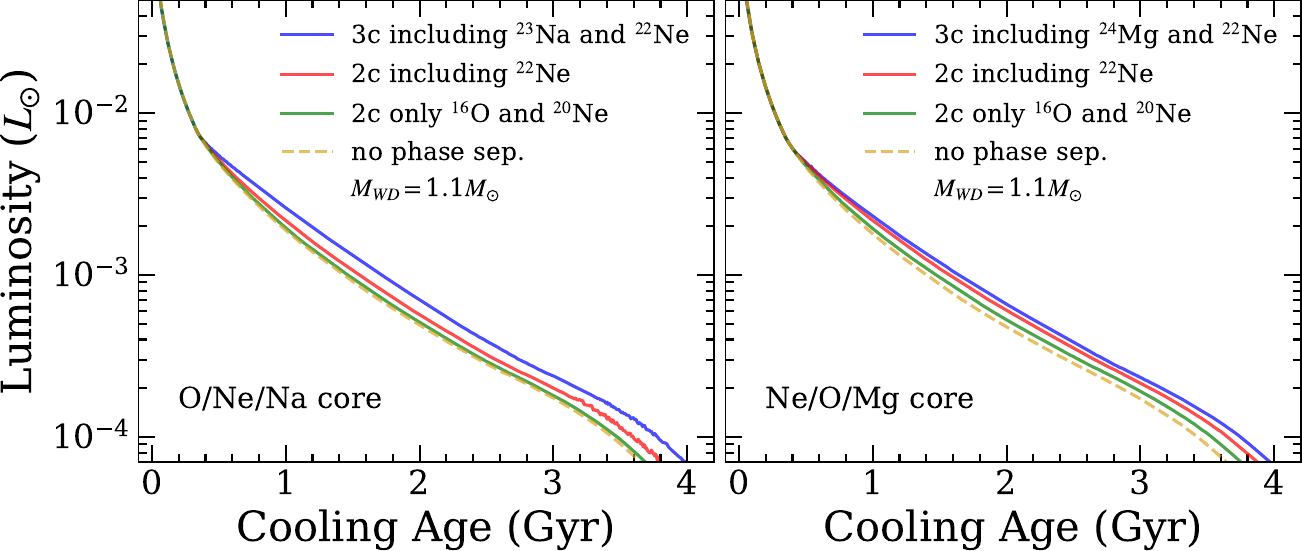}
  \caption{Cooling curves of the $1.1\ M_{\odot}$ O/Ne UMWDs models used in this work. The left panel shows O/Ne/Na core cases, while the right panels show the Ne/O/Mg cores. 3c stands for 3-component, while 2c stands for 2-component to differentiate the phase separation treatment. 
  }
    \label{fig:cool}
\end{figure*}

Determining the cooling age of massive WDs also remains an open problem. We use our 3-species phase separation to investigate the amount of cooling delay provided by the fractionation process in UMWDs. 
\begin{figure}
    \centering
    \includegraphics[width=0.47\textwidth]{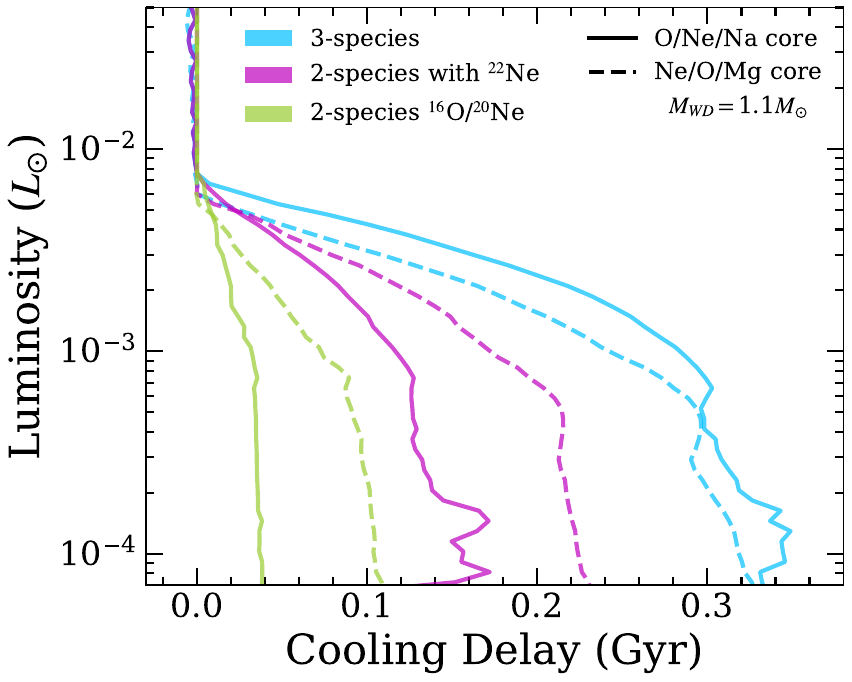}
  \caption{Cooling delay induced at a given luminosity of the $1.1\ M_{\odot}$ O/Ne UMWDs models used in this work. The delays are computed as the difference in the age of the labeled model and the model without phase separation at fixed luminosity. 3c stands for 3-component, while 2c stands for 2-component to differentiate the phase separation treatment.
  }
    \label{fig:delay}
\end{figure}

\begin{figure*}
    \centering
    \includegraphics[width=0.99\textwidth]{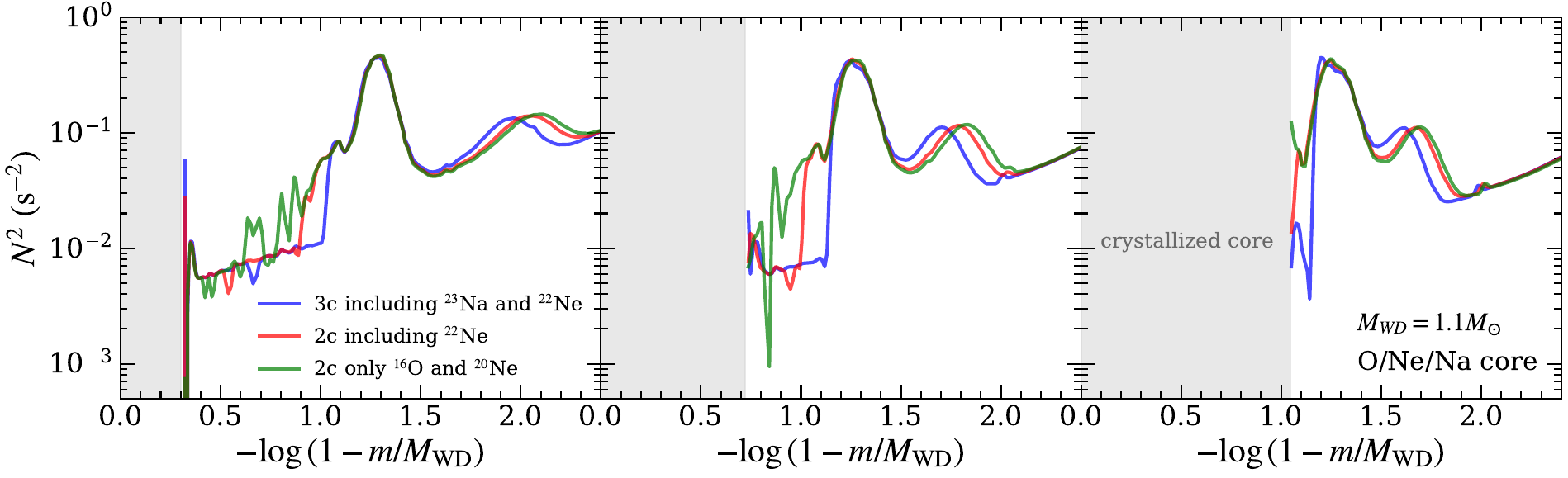}
  \caption{Brunt-V\"ais\"al\"a frequency as a function of the mass coordinate $-\log{(1-m/M_{\mathrm{WD}})}$ at three different stages of crystallization for $1.1\ M_{\odot}$ O/Ne/Na WD models. 3c stands for 3-component, while 2c stands for 2-component to differentiate the phase separation treatment.}
    \label{fig:brunt}
\end{figure*}

In Figure \ref{fig:cool} we show the cooling curve of a $1.1\ M_{\odot}$ WD for the different cases of the phase separation investigated in the current work compared to the case of cooling without phase separation, but including the latent heat term. On the left panel, we show the case of an O/Ne/Na core, while on the right panel, we show the case of Ne/O/Mg internal composition. We see that the cooling curve without fractionation and the 2-species case are marginally similar for the O/Ne/Na core with an almost imperceptible delay, while adding only the $\mathrm{^{22}Ne}$ does not seem to add a significant delay either, but the curves are now distinguishable. However, the case of 3-species separation with $\mathrm{^{23}Na}$ generates a very different cooling curve from the other three cases for the O/Ne/Na core. For the case of the Ne/O/Mg core, each additional component to the crystallization modeling seems to add a similar amount of cooling delay. For instance, the 3-component separation with $\mathrm{^{24}Mg}$ curve is separated in luminosity with the 2-component with $\mathrm{^{22}Ne}$ by the same amount that the simplest 2-component when compared with the curve without fractionation.

We computed the amount of cooling delay observed at each luminosity when comparing the cooling curves that include fractionation with the case without phase separation. In Figure \ref{fig:delay} we show these delays. As mentioned before, for the O/Ne/Na core, we see that the longest delay is given by the full implementation of the 3-components separation, providing a delay of up to $\sim 0.35$ Gyr, which is about 2.5 times the delay of the 2-species with $\mathrm{^{22}Ne}$ and almost 10 times the delay provided by the simplest case of phase separation. For the Ne/O/Mg core, the amount of delay provided by each degree of complexity of the phase separation implementation is about $\gtrsim 0.11$ Gyr, which translates into the 3-species providing a $\sim 0.33$ Gyr delay.

Additionally, we estimated the magnitude of the total gravitational energy released by element redistribution from our $1.1\ \mathrm{M_{\odot}}$ models with O/Ne/Na cores. We obtained $E_{\mathrm{grav,tot,2c}}\approx6.6\times{10^{46}}\ \mathrm{erg}$ and  $E_{\mathrm{grav,tot,2c-^{22}Ne}}\approx2.9\times{10^{47}}\ \mathrm{erg}$, for the 2-species phase separation without and with $\mathrm{^{22}Ne}$. For the 3-species case, we got 
$E_{\mathrm{grav,tot,3c}}\approx8.4\times{10^{47}}\ \mathrm{erg}$. Taking the luminosity when crystallization starts $\log{(L_{\mathrm{crys}}/L_{\odot})}\approx -2.1$, we obtain an approximate delay of $E_{\mathrm{grav,tot,2c}}/L_{\mathrm{crys}}\approx0.07$ Gyr, $E_{\mathrm{grav,tot,2c-^{22}Ne}}/L_{\mathrm{crys}}\approx0.311$ Gyr, and $E_{\mathrm{grav,tot,3c}}/L_{\mathrm{crys}}\approx 0.89$ Gyr, which are of the same order of magnitude as the results shown in Figure \ref{fig:delay}.

These delays are much smaller to those needed for explaining the problem of the extreme cooling delay observed in massive WDs \citep{Cheng2019}. The theory that best addresses such amount of delay is the distillation driven by poor-$\mathrm{^{22}Ne}$ crystals predicted to form in C/O/Ne mixtures, where $X_{\mathrm{^{22}Ne}}\lesssim0.05$ \citep{Blouin2021}. Some works \citep{Blouin2021, Bedard2024} have implemented distillation using the C/O/Ne phase diagrams, but they only covered a small range of abundances in the ternary space of parameters, restricting the abundances of Ne to only a small fraction. In addition, they only focused on investigating the energetics of this process to explain the cooling delay and have not addressed in detail the buoyant motions associated with it. While we have not investigated or implemented the distillation process in MESA, it would be interesting to extend the methods developed here to this case. In Appendix \ref{appB}, we compare our C/O/Ne diagram in the relevant parameter space for distillation with the one used in \citet{Blouin2021}.

\subsection{Implications for Pulsating UMWD Models}

In pulsating UMWDs, photometric variations can be associated with long-period gravity($g-$)modes \citep[see][for a review]{Althaus2010}. One common assumption is that these pulsations are adiabatic, and they do not propagate inside the solid lattice \citep{MontgomeryWinget1999}. The properties of the $g-$modes depend on the Brunt-V\"ais\"al\"a frequency
\begin{equation}\label{eq:brunt}
N^{2}=\frac{g^{2}\rho}{P}\frac{\chi_{T}}{\chi_{\rho}}{}(\nabla_{\mathrm{ad}}-\nabla-\nabla_{\mathrm{com}}).
\end{equation}
In Figure \ref{fig:brunt}, we show the Brunt-V\"ais\"al\"a frequency profile for our 1.1 $M_{\odot}$ O/Ne/Na WD models. We show three stages of the crystallization: $52\%$, $81\%$, and $91\%$ of the mass solidified. Moreover, the three different versions of the phase separation analyzed in the current work are compared for each case. We note that $N^{2}$ is flatter in the liquid regions just outside the solid core as more components are added to the phase transition prescription.

For the cases of $81\%$ and $91\%$ of the mass crystallized in Figure \ref{fig:brunt}, we see that the difference in $N^{2}$ just above the crystal core is up to about one order of magnitude when comparing $\mathrm{^{16}O}/\mathrm{^{20}Ne}$ and $\mathrm{^{16}O}/(\mathrm{^{20}Ne}+\mathrm{^{22}Ne})/\mathrm{^{23}Na}$ phase separation. This feature is a direct effect of the more extended mixing when considering more species in crystallization (as shown in Figure \ref{fig:ONeNa_sep}), which reduces the value of the composition gradient $-\nabla_{\mathrm{com}}$ in the calculation of $N^{2}$. Note, for example, in the middle panel of Figure \ref{fig:brunt}, when taking 3- species instead of 2-species separation, $N^{2}$ is reduced from $\sim 10^{-1}\ \mathrm{s^{-2}}$ to $\lesssim 10^{-2}\ \mathrm{s^{-2}}$ up to a zone corresponding to $\gtrsim91\%$ of the mass coordinate. The reduced $N^{2}$ in the mixed zone in the 3-species case, where the oscillations propagate slowly, could mimic a larger crystallized zone from a model only with 2-species separation, where the pulsations would not propagate in this zone\footnote{This is similar to the statement of \citet{MontgomeryDunlap2024} about considering a neutrally buoyant profile $N^{2}=0$ in the liquid layers outside solid core where thermohaline mixing should occur.}. Thus, having only a 2-species phase separation model can lead to a wrong prediction of the solid core size. The core size prediction is a key quantity for unveiling the mass of the WD and the core composition, as the crystallization process depends on both features \citep{Camisassa2019, Camisassa2022a, Camisassa2022b}.

\section{Summary and Discussion}\label{sec:sum_disc}
We have implemented a 3-component phase separation into the stellar evolution code MESA to study the cooling and evolution of UMWDs. We constructed UMWD models with O/Ne/Na and Ne/O/Mg cores to test our modified phase separation routine, and compared it with the current 2-species phase separation implemented in MESA. We noted that species such as $\mathrm{^{23}Na}$ and $\mathrm{^{24}Mg}$ can have a great impact in the compositional profile of the WD as crystallization advances, even when these elements make up only $\lesssim10\%$ of the total core composition of O/Ne WDs. The compositional change is not only induced by the fractionation of the elements at the solid-liquid interface, but also because of the additional mixing induced in the liquid phase, which occurs due to the heavier elements being preferentially retained in the solid phase. This is similar to the 2-species separation, but now including the third element, which is the heaviest and in most cases follows the Ne-enrichment in the solid.

In addition to the implementation of the 3-species phase separation, we investigated the case of 2-species separation but now including the $\mathrm{^{22}Ne}$ isotope, as current stellar evolution codes only account for $\mathrm{^{16}O}/\mathrm{^{20}Ne}$ separation when studying UMWDs \citep[e.g.][]{Camisassa2019}. We showed that despite its quite small presence in the core, $\lesssim2\%$, $\mathrm{^{22}Ne}$ has a significant impact on the composition profile of the WD as it crystallizes. For instance, the mixing is much more extended in the liquid phase, and towards more than a half of the core crystallized, the 2-species separation including $\mathrm{^{22}Ne}$ resembles the chemical profiles of $\mathrm{^{16}O}$ and  $\mathrm{^{20}Ne}$ of the 3-species separation case, in both, liquid and solid phase (Figures \ref{fig:ONeNa_sep} and \ref{fig:ONeMg_sep}).

We used the 3-species phase separation to study the effect on UMWD models. We first showed that the convective efficiency induced by the compositional gradients in the liquid phase upon crystallization is increased when adding more species to the mixture considered for fractionation. To measure the strength of compositional convection, we used a general form for $n$-species of the dimensionless $\tau$ parameter used in \citet{Castro-Tapia2024}, where $\tau>1$ implies fast overturning convection, and $0<\tau<1$ implies inefficient (thermohaline) convection. Similar to \citet{Castro-Tapia2024}, we observed both convection regimes, fast first, and after a short time, the slow regime for most of the evolution. However, we show that the increment in the value of $\tau$ for 3-species separation also makes the fast regime longer by up to 2 orders of magnitude compared to the current MESA version of the 2-species phase separation.

We also compared the $\mathrm{^{16}O}$/$\mathrm{^{20}Ne}$ phase separation with the case of 2-species with $\mathrm{^{22}Ne}$. Interestingly, this small change in the implementation has a similar effect on the convective efficiency as the 3-species phase separation, with both cases being almost equal when taking an Ne/O/Mg core. Since the species $\mathrm{^{16}O}$, $\mathrm{^{20}Ne}$, and $\mathrm{^{24}Mg}$ have an equal number of neutrons and protons, we inferred that neutron-rich species such as $\mathrm{^{22}Ne}$ and $\mathrm{^{23}Na}$ have a greater impact on the convective efficiency, even when they account for small abundances.

The convective efficiency is key to studying the theory of the crystallization-driven dynamo for explaining magnetic WDs \citep{Isern2017}. While this theory has been studied for UMWDs, these studies have focused on the timing of the crystallization onset and appearance of the magnetic field at the surface when assuming O/Ne or C/O cores \citep{Camisassa2022b, BlatmanGinzburg2024}. Furthermore, while these studies favor the early breakout of the field on the surface for O/Ne mixtures, they have not addressed whether convection is vigorous enough to, in principle, generate the magnetic field. An analysis similar to \citet{Castro-Tapia2024b} is necessary to fully assess the strength and evolution of the magnetic field for this theory, but applied to O/Ne UMWDs, since in that work, only non-ultramassive C/O WDs are analyzed. Our current study paves the path to analyze the magnetic field generation on O/Ne UMWDs by giving more accurate estimations (than those that can be obtained with current public cooling models) on the timing of efficient convection and buoyancy parameters ($\tau$) upon crystallization. 

We computed the cooling curves for the models with different versions of the phase separation studied in this work and demonstrated that the 3-species implementation must be used to constrain the age of UMWDs accurately. We computed a delay of $\sim 0.34$ Gyr to up to $\sim 0.89$ Gyr for WDs with O/Ne/Na cores, which is about 10 times larger than what we predict for the case of the current 2-species phase transition routine in MESA ($\sim 0.04$-$0.07$ Gyr). For the case of O/Ne/Mg, the difference, although less dramatic, is a delay of 3 times larger, being $\sim 0.33$ Gyr for the 3-species case and $\sim 0.1$ for the simplest 2-species separation. It is worth noting that our models have a pure He atmosphere, which usually results in quicker cooling than in the case of H-rich atmospheres \citep{Camisassa2019}. Then, it would be interesting to see how different the cooling delay introduced by a 3-species phase separation is for O/Ne UMWDs with H-rich atmospheres.

While the delay we observed with our implementation is much smaller than the one expected to explain the extreme cooling delay observed in some massive WDs \citep{Cheng2019}, the quantified delays are a significant fraction of the age of O/Ne UMWDs ($\sim 10\%$). As proposed by several studies \citep{Blouin2021, Shen2023, Bedard2024}, multi-Gyr UMWDs are likely to be formed with C/O cores in merger events, which leave them with impurities of $\mathrm{^{22}Ne}$, $\mathrm{^{25}Mg}$, and $\mathrm{^{26}Mg}$, that as predicted by the C/O/Ne and C/O/Mg phase diagrams, may drive distillation for certain range of abundances. In this paper, we have assumed that the solid crystals that form have a larger density than the surrounding liquid, so that we are not able to treat distillation. However, our phase separation extension includes the C/O/Ne and C/O/Mg diagrams, and can be used to test this theory in MESA, although considering a tight range of abundances due to the small abundance contrasts that drive distillation and the differences in the phase diagrams computed from diverse methods (Appendix \ref{appB}).

We proposed that, by implementing our 3-species phase separation extension, the predictions on stellar parameters made for pulsating UMWDs \citep[e.g.,][]{Corsico2019} can be improved. Given that the addition of more species to the fractionation process enhances the mixing outside the solid phase, we predict that the frequency at which gravity modes may propagate in the mixed zone will be reduced by at least one order of magnitude, being the smallest frequency in the propagation zone. Given these long periods and the fact that the gravity modes do not propagate in the solid phase, models with only a 2-species separation may not be predicting the correct solid core size. It will be interesting to explore this in future work.

We have quantified the effect of including a third element on O/Ne UMWD models using our extension of the 3-species phase separation in the stellar evolution code MESA. More work on UMWDs is needed to explore the potential of these models. Additionally, more advanced techniques must be applied to improve the precision of the diagrams used in this work, since the implementation of distillation in stellar evolution codes is the special interest to address the extreme cooling delay in the so-called Q-branch in the color-magnitude diagram. Finally, the 3-species phase separation implementation we presented could be used to explore the phase transition in C/O WDs to see if impurities can affect the crystallization-driven convection, when distillation does not occur.

\begin{acknowledgements}
 We thank Mike Montgomery, Bart Dunlap, Maria Camisassa, Evan Bauer, Claudia Aguilera-Gómez, Alejandro Córsico, Leandro Althaus, and Stefano Bagnulo for useful discussions during the 23rd European Workshop on White Dwarfs. We thank Simon Blouin for providing the C/O/Ne diagram from \citet{Blouin2021}. M.C.-T. thanks J.R. Fuentes, for his hospitality during a visit to CU Boulder. M.C.-T. is supported by the Fonds de recherche du Québec - Nature et technologies through a doctoral scholarship (\href{	https://doi.org/10.69777/366094}{DOI:10.69777/366094}). We acknowledge support by NSERC Discovery Grant RGPIN-2023-03620. A.C. and M.C.-T. are members of the Centre de Recherche en Astrophysique du Québec (CRAQ). 
\end{acknowledgements}

\appendix
\section{Correction to solid core coordinate in MESA}\label{appA}
\begin{figure}
    \centering
    \includegraphics[width=0.49\textwidth]{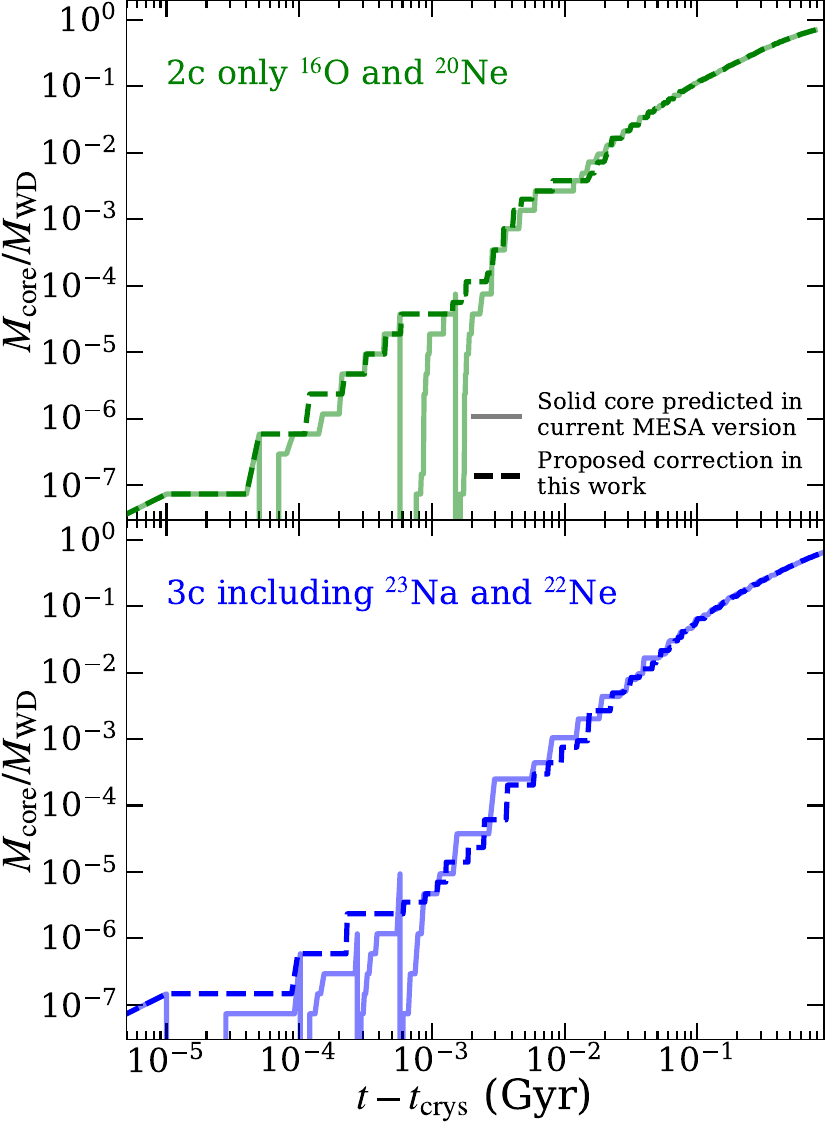}
    \includegraphics[width=0.49\textwidth]{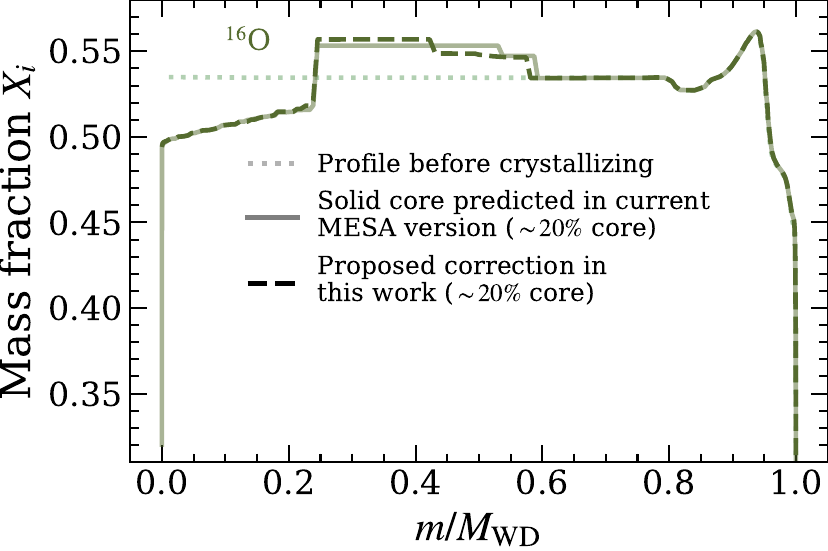}
  \caption{The upper panels show the growth of the solid core as a function of the time since crystallization started for different implementations of the phase separation routine in a $1.1\ M_{\odot}$ O/Ne/Na WD model. We show cases without and with our proposed correction to the current phase separation routine in MESA for identifying the location of the solid core (see text for more details). The lower panel shows the $\mathrm{^{16}O}$ composition profile of the model when applying the 2-species phase separation without and with the proposed correction.  
  }
    \label{fig:core_corr}
\end{figure}

As discussed in Section \ref{sec:2.2}, in the current version of MESA, the implementation of the phase separation fails to correctly follow the mass of the crystal core. This problem arises at the small timesteps required to follow the initial growth of the core, for which the energy deposited in the central cells does not have time to diffuse out, resulting in periodic melting of the core. Our proposed correction to the phase separation routine is that once the core is $>0$, it remains as such, and allows the extra energy to diffuse without pushing the core size to zero again. In this Appendix, we compare the old and new methods in more detail.

In the upper panels of Figure \ref{fig:core_corr} we show in logarithmic scale the size in mass of the crystallized core predicted from the current implementation in MESA for the $1.1\ M_{\odot}$ O/Ne/Na WD model compared to the version applying our small correction to the code. We show two versions of the elements redistribution, the current $\mathrm{^{16}O}/\mathrm{^{20}Ne}$ separation and the 3-component version studied through this paper. We note how the core size is pushed to 0 in the current MESA version. In the cases shown, we started limiting the timestep to $<10^{4}$ yr, increasing up to $<10^{6}$ yr, and then without restriction at $t-t_{\mathrm{crys}}\gtrsim 10^{-3}$ Gyr. When we applied our correction, we noted that the timestep size did not unexpectedly reduce the core size.

The reduction of the core size in the current MESA phase separation routine artificially melts the core for some time while the energy is diffused out. This implies a problem when fractionating the abundances in the solid and liquid phases. Since the methods used to compute the abundances in both states only consider the transition from liquid to solid, the abundances in the central cells will change dramatically as the core is melted and frozen again, only considering fractionation in one direction. In the lower panel of Figure \ref{fig:core_corr}, we show the abundance of $\mathrm{^{16}O}$ for two stages, before and after the onset the crystallization for the $1.1\ M_{\odot}$ O/Ne/Na WD model, and considering the simplest 2-component phase separation. We note that the central abundance for the current MESA implementation is significantly different from the rest of the crystallized part, which is a direct consequence of the artificially melted core. In contrast, our correction shows a more continuous change in abundance, as expected. Additionally, we note that the abundance profile is also modified in the liquid phases as the phase transition is applied multiple times in the core for the current MESA routine.

\section{Comparison of Phase Diagrams for Distillation Modeling}\label{appB}
\begin{figure}
    \centering
    \includegraphics[width=0.475\textwidth]{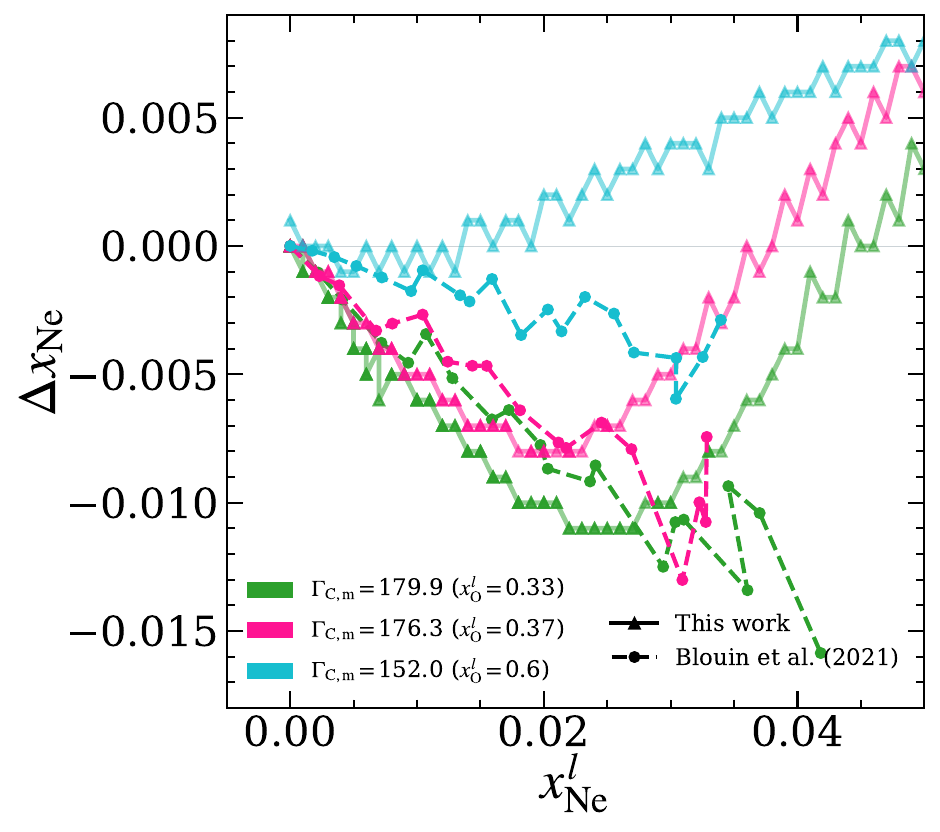}
  \caption{Three cases of $\Delta{x_{\mathrm{Ne}}}$ as a function of $x^{l}_{\mathrm{Ne}}$ at fixed $\Gamma_{\mathrm{C,m}}$ obtained from our C/O/Ne diagram and the one from \citet{Blouin2021}. The values in parenthesis for each $\Gamma_{\mathrm{C,m}}$ is the oxygen concentration at $x^{l}_{\mathrm{Ne}}=0$.
  }
    \label{fig:delta_Ne_b}
\end{figure}

We compare our C/O/Ne phase diagram with the results of \citet{Blouin2021} to see if our diagrams can be used to study the distillation of Ne-poor crystals in the interior of massive C/O WDs that probably evolved from merger processes. We consider that our diagram predicts the concentration differences with a precision of $0.001$ since this is the separation considered when making our interpolation tables. 

For a comparison between our C/O/Ne diagram and the one in \citet{Blouin2021}, in Figure \ref{fig:delta_Ne_b}, we show three cases of $\Delta{x_{\mathrm{Ne}}}$ as a function of $x^{l}_{\mathrm{Ne}}$ at fixed $\Gamma_{\mathrm{C,m}}$ with different oxygen concentrations at $x^{l}_{\mathrm{Ne}}=0$. We note that, for large values of $\Gamma_{\mathrm{C,m}}$, both diagrams coincide very well. However, we note that for all the values of $x^{l}_{\mathrm{Ne}}$ shown the diagram of \citet{Blouin2021} predict a $\Delta{x_{\mathrm{Ne}}}\leq0$ which decreases with $x^{l}_{\mathrm{Ne}}$ regardless of $\Gamma_{\mathrm{C,m}}$, whereas our diagrams predict that at certain point $\Delta{x_{\mathrm{Ne}}}$ increases reaching values $>0$ in the range shown. Such behavior is also expected in the diagrams of \citet{Blouin2021}, but for larger values of $x^{l}_{\mathrm{Ne}}$. For example, their Figure 1 shows that for the $\Gamma_{\mathrm{C,m}}=179.8$ case, $\Delta{x_{\mathrm{Ne}}}$ must increase at  $x^{l}_{\mathrm{Ne}}\approx 0.1$, and becomes positive at $x^{l}_{\mathrm{Ne}}\approx 0.15$. These differences in the exact abundance of Ne, where $\Delta{x_{\mathrm{Ne}}}$ changes its trend, could be related to the different methodologies used for computing the phase diagrams. The difference become more dramatic for $\Gamma_{\mathrm{C,m}}=152$, where our prediction shows that for very small $x^{l}_{\mathrm{Ne}}$ values, $\Delta{x_{\mathrm{Ne}}}$ becomes positive.

Given these results, and the fact that distillation can only occur when $\Delta{x_{\mathrm{Ne}}}<0$, this process in WDs with C/O cores and $\mathrm{^{22}Ne}$ impurities may be studied using our methods and diagrams, but only for a certain range of abundances and coupling parameters, mostly $x^{l}_{\mathrm{Ne}}\lesssim 0.05$ and $\Gamma_{\mathrm{C,m}}\gtrsim170$. This range of parameters makes more plausible to study with our methods the case described by \citet{Blouin2021} where C/O crystallization may occur in the core without driving distillation, and as $\Gamma_{\mathrm{C,m}}$ increases and $x^{l}_{\mathrm{O}}$ decreases due to fractionation and O-enrichment in the liquid phase, distillation can be driven in a shell out of the crystal core forming a Ne-rich liquid shell \citep[see Figure 3, case (c) in][for more details]{Blouin2021}.



\begin{thebibliography}{}
\expandafter\ifx\csname natexlab\endcsname\relax\def\natexlab#1{#1}\fi

\bibitem[{{Althaus} {et~al.}(2010){Althaus}, {C{\'o}rsico}, {Isern}, \& {Garc{\'\i}a-Berro}}]{Althaus2010}
{Althaus}, L.~G., {C{\'o}rsico}, A.~H., {Isern}, J., \& {Garc{\'\i}a-Berro}, E. 2010, \aapr, 18, 471

\bibitem[{{Althaus} {et~al.}(2012){Althaus}, {Garc{\'\i}a-Berro}, {Isern}, {C{\'o}rsico}, \& {Miller Bertolami}}]{Althaus2012}
{Althaus}, L.~G., {Garc{\'\i}a-Berro}, E., {Isern}, J., {C{\'o}rsico}, A.~H., \& {Miller Bertolami}, M.~M. 2012, \aap, 537, A33

\bibitem[{{Bauer}(2023)}]{Bauer2023}
{Bauer}, E.~B. 2023, \apj, 950, 115

\bibitem[{{Bauer} {et~al.}(2020){Bauer}, {Schwab}, {Bildsten}, \& {Cheng}}]{Bauer2020}
{Bauer}, E.~B., {Schwab}, J., {Bildsten}, L., \& {Cheng}, S. 2020, \apj, 902, 93

\bibitem[{{B{\'e}dard} {et~al.}(2024){B{\'e}dard}, {Blouin}, \& {Cheng}}]{Bedard2024}
{B{\'e}dard}, A., {Blouin}, S., \& {Cheng}, S. 2024, \nat, 627, 286

\bibitem[{{Bildsten} \& {Hall}(2001)}]{BildstenHall2001}
{Bildsten}, L., \& {Hall}, D.~M. 2001, \apjl, 549, L219

\bibitem[{{Blatman} \& {Ginzburg}(2024)}]{BlatmanGinzburg2024}
{Blatman}, D., \& {Ginzburg}, S. 2024, \mnras, 528, 3153

\bibitem[{{Blouin} \& {Daligault}(2021{\natexlab{a}})}]{Blouin2021b}
{Blouin}, S., \& {Daligault}, J. 2021{\natexlab{a}}, \pre, 103, 043204

\bibitem[{{Blouin} \& {Daligault}(2021{\natexlab{b}})}]{Blouin2021c}
---. 2021{\natexlab{b}}, \apj, 919, 87

\bibitem[{{Blouin} {et~al.}(2021){Blouin}, {Daligault}, \& {Saumon}}]{Blouin2021}
{Blouin}, S., {Daligault}, J., \& {Saumon}, D. 2021, \apjl, 911, L5

\bibitem[{{Blouin} {et~al.}(2020){Blouin}, {Daligault}, {Saumon}, {B{\'e}dard}, \& {Brassard}}]{Blouin2020}
{Blouin}, S., {Daligault}, J., {Saumon}, D., {B{\'e}dard}, A., \& {Brassard}, P. 2020, \aap, 640, L11

\bibitem[{{Calcaferro} {et~al.}(2024){Calcaferro}, {C{\'o}rsico}, {Uzundag}, {Althaus}, {Kepler}, \& {Werner}}]{Calcaferro2024}
{Calcaferro}, L.~M., {C{\'o}rsico}, A.~H., {Uzundag}, M., {et~al.} 2024, \aap, 691, A194

\bibitem[{{Camisassa} {et~al.}(2022{\natexlab{a}}){Camisassa}, {Althaus}, {Koester}, {Torres}, {Gil-Pons}, \& {C{\'o}rsico}}]{Camisassa2022a}
{Camisassa}, M.~E., {Althaus}, L.~G., {Koester}, D., {et~al.} 2022{\natexlab{a}}, \mnras, 511, 5198

\bibitem[{{Camisassa} {et~al.}(2022{\natexlab{b}}){Camisassa}, {Raddi}, {Althaus}, {Isern}, {Rebassa-Mansergas}, {Torres}, {C{\'o}rsico}, \& {Korre}}]{Camisassa2022b}
{Camisassa}, M.~E., {Raddi}, R., {Althaus}, L.~G., {et~al.} 2022{\natexlab{b}}, \mnras, 516, L1

\bibitem[{{Camisassa} {et~al.}(2019){Camisassa}, {Althaus}, {C{\'o}rsico}, {De Ger{\'o}nimo}, {Miller Bertolami}, {Novarino}, {Rohrmann}, {Wachlin}, \& {Garc{\'\i}a-Berro}}]{Camisassa2019}
{Camisassa}, M.~E., {Althaus}, L.~G., {C{\'o}rsico}, A.~H., {et~al.} 2019, \aap, 625, A87

\bibitem[{{Caplan} {et~al.}(2018){Caplan}, {Cumming}, {Berry}, {Horowitz}, \& {Mckinven}}]{Caplan2018}
{Caplan}, M.~E., {Cumming}, A., {Berry}, D.~K., {Horowitz}, C.~J., \& {Mckinven}, R. 2018, \apj, 860, 148

\bibitem[{{Caplan} {et~al.}(2021){Caplan}, {Freeman}, {Horowitz}, {Cumming}, \& {Bellinger}}]{Caplan2021}
{Caplan}, M.~E., {Freeman}, I.~F., {Horowitz}, C.~J., {Cumming}, A., \& {Bellinger}, E.~P. 2021, \apjl, 919, L12

\bibitem[{{Caplan} {et~al.}(2020){Caplan}, {Horowitz}, \& {Cumming}}]{Caplan2020}
{Caplan}, M.~E., {Horowitz}, C.~J., \& {Cumming}, A. 2020, \apjl, 902, L44

\bibitem[{{Castro-Tapia} {et~al.}(2024{\natexlab{a}}){Castro-Tapia}, {Cumming}, \& {Fuentes}}]{Castro-Tapia2024}
{Castro-Tapia}, M., {Cumming}, A., \& {Fuentes}, J.~R. 2024{\natexlab{a}}, \apj, 969, 10

\bibitem[{{Castro-Tapia} {et~al.}(2024{\natexlab{b}}){Castro-Tapia}, {Zhang}, \& {Cumming}}]{Castro-Tapia2024b}
{Castro-Tapia}, M., {Zhang}, S., \& {Cumming}, A. 2024{\natexlab{b}}, \apj, 975, 63

\bibitem[{{Cheng} {et~al.}(2019){Cheng}, {Cummings}, \& {M{\'e}nard}}]{Cheng2019}
{Cheng}, S., {Cummings}, J.~D., \& {M{\'e}nard}, B. 2019, \apj, 886, 100

\bibitem[{{Cheng} {et~al.}(2020){Cheng}, {Cummings}, {M{\'e}nard}, \& {Toonen}}]{Cheng2020}
{Cheng}, S., {Cummings}, J.~D., {M{\'e}nard}, B., \& {Toonen}, S. 2020, \apj, 891, 160

\bibitem[{{C{\'o}rsico} {et~al.}(2019){C{\'o}rsico}, {De Ger{\'o}nimo}, {Camisassa}, \& {Althaus}}]{Corsico2019}
{C{\'o}rsico}, A.~H., {De Ger{\'o}nimo}, F.~C., {Camisassa}, M.~E., \& {Althaus}, L.~G. 2019, \aap, 632, A119

\bibitem[{{De Ger{\'o}nimo} {et~al.}(2025){De Ger{\'o}nimo}, {Uzundag}, {Rebassa-Mansergas}, {Brown}, {Kilic}, {C{\'o}rsico}, {Jewett}, \& {Moss}}]{DeGeronimo2025}
{De Ger{\'o}nimo}, F.~C., {Uzundag}, M., {Rebassa-Mansergas}, A., {et~al.} 2025, \apjl, 980, L9

\bibitem[{{Fuentes} {et~al.}(2024){Fuentes}, {Castro-Tapia}, \& {Cumming}}]{Fuentes2024}
{Fuentes}, J.~R., {Castro-Tapia}, M., \& {Cumming}, A. 2024, \apjl, 964, L15

\bibitem[{{Fuentes} {et~al.}(2023){Fuentes}, {Cumming}, {Castro-Tapia}, \& {Anders}}]{Fuentes2023}
{Fuentes}, J.~R., {Cumming}, A., {Castro-Tapia}, M., \& {Anders}, E.~H. 2023, \apj, 950, 73

\bibitem[{{Ginzburg} {et~al.}(2022){Ginzburg}, {Fuller}, {Kawka}, \& {Caiazzo}}]{Ginzburg2022}
{Ginzburg}, S., {Fuller}, J., {Kawka}, A., \& {Caiazzo}, I. 2022, \mnras, 514, 4111

\bibitem[{{Horowitz} {et~al.}(2010){Horowitz}, {Schneider}, \& {Berry}}]{Horowitz2010}
{Horowitz}, C.~J., {Schneider}, A.~S., \& {Berry}, D.~K. 2010, \prl, 104, 231101

\bibitem[{{Isern} {et~al.}(2000){Isern}, {Garc{\'\i}a-Berro}, {Hernanz}, \& {Chabrier}}]{Isern2000}
{Isern}, J., {Garc{\'\i}a-Berro}, E., {Hernanz}, M., \& {Chabrier}, G. 2000, \apj, 528, 397

\bibitem[{{Isern} {et~al.}(2017){Isern}, {Garc{\'\i}a-Berro}, {K{\"u}lebi}, \& {Lor{\'e}n-Aguilar}}]{Isern2017}
{Isern}, J., {Garc{\'\i}a-Berro}, E., {K{\"u}lebi}, B., \& {Lor{\'e}n-Aguilar}, P. 2017, \apjl, 836, L28

\bibitem[{{Isern} {et~al.}(1991){Isern}, {Hernanz}, {Mochkovitch}, \& {Garcia-Berro}}]{Isern1991}
{Isern}, J., {Hernanz}, M., {Mochkovitch}, R., \& {Garcia-Berro}, E. 1991, \aap, 241, L29

\bibitem[{{Isern} {et~al.}(1997){Isern}, {Mochkovitch}, {Garc{\'\i}a-Berro}, \& {Hernanz}}]{Isern1997}
{Isern}, J., {Mochkovitch}, R., {Garc{\'\i}a-Berro}, E., \& {Hernanz}, M. 1997, \apj, 485, 308

\bibitem[{{Jermyn} {et~al.}(2021){Jermyn}, {Schwab}, {Bauer}, {Timmes}, \& {Potekhin}}]{Jermyn2021}
{Jermyn}, A.~S., {Schwab}, J., {Bauer}, E., {Timmes}, F.~X., \& {Potekhin}, A.~Y. 2021, \apj, 913, 72

\bibitem[{{Jermyn} {et~al.}(2023){Jermyn}, {Bauer}, {Schwab}, {Farmer}, {Ball}, {Bellinger}, {Dotter}, {Joyce}, {Marchant}, {Mombarg}, {Wolf}, {Sunny Wong}, {Cinquegrana}, {Farrell}, {Smolec}, {Thoul}, {Cantiello}, {Herwig}, {Toloza}, {Bildsten}, {Townsend}, \& {Timmes}}]{Jermyn2023}
{Jermyn}, A.~S., {Bauer}, E.~B., {Schwab}, J., {et~al.} 2023, \apjs, 265, 15

\bibitem[{{Kilic} {et~al.}(2023){Kilic}, {C{\'o}rsico}, {Moss}, {Jewett}, {De Ger{\'o}nimo}, \& {Althaus}}]{Kilic2023}
{Kilic}, M., {C{\'o}rsico}, A.~H., {Moss}, A.~G., {et~al.} 2023, \mnras, 522, 2181

\bibitem[{{Lauffer} {et~al.}(2018){Lauffer}, {Romero}, \& {Kepler}}]{Lauffer2018}
{Lauffer}, G.~R., {Romero}, A.~D., \& {Kepler}, S.~O. 2018, \mnras, 480, 1547

\bibitem[{{Medin} \& {Cumming}(2010)}]{MedinCumming2010}
{Medin}, Z., \& {Cumming}, A. 2010, \pre, 81, 036107

\bibitem[{{Medin} \& {Cumming}(2015)}]{MedinCumming2015}
---. 2015, \apj, 802, 29

\bibitem[{{Mochkovitch}(1983)}]{Mochkovitch1983}
{Mochkovitch}, R. 1983, \aap, 122, 212

\bibitem[{{Montgomery} \& {Dunlap}(2024)}]{MontgomeryDunlap2024}
{Montgomery}, M.~H., \& {Dunlap}, B.~H. 2024, \apj, 961, 197

\bibitem[{{Montgomery} \& {Winget}(1999)}]{MontgomeryWinget1999}
{Montgomery}, M.~H., \& {Winget}, D.~E. 1999, \apj, 526, 976

\bibitem[{{Nandkumar} \& {Pethick}(1984)}]{NandkumarPethick1984}
{Nandkumar}, R., \& {Pethick}, C.~J. 1984, \mnras, 209, 511

\bibitem[{{Paxton} {et~al.}(2011){Paxton}, {Bildsten}, {Dotter}, {Herwig}, {Lesaffre}, \& {Timmes}}]{Paxton2011}
{Paxton}, B., {Bildsten}, L., {Dotter}, A., {et~al.} 2011, \apjs, 192, 3

\bibitem[{{Paxton} {et~al.}(2013){Paxton}, {Cantiello}, {Arras}, {Bildsten}, {Brown}, {Dotter}, {Mankovich}, {Montgomery}, {Stello}, {Timmes}, \& {Townsend}}]{Paxton2013}
{Paxton}, B., {Cantiello}, M., {Arras}, P., {et~al.} 2013, \apjs, 208, 4

\bibitem[{{Paxton} {et~al.}(2015){Paxton}, {Marchant}, {Schwab}, {Bauer}, {Bildsten}, {Cantiello}, {Dessart}, {Farmer}, {Hu}, {Langer}, {Townsend}, {Townsley}, \& {Timmes}}]{Paxton2015}
{Paxton}, B., {Marchant}, P., {Schwab}, J., {et~al.} 2015, \apjs, 220, 15

\bibitem[{{Paxton} {et~al.}(2018){Paxton}, {Schwab}, {Bauer}, {Bildsten}, {Blinnikov}, {Duffell}, {Farmer}, {Goldberg}, {Marchant}, {Sorokina}, {Thoul}, {Townsend}, \& {Timmes}}]{Paxton2018}
{Paxton}, B., {Schwab}, J., {Bauer}, E.~B., {et~al.} 2018, \apjs, 234, 34

\bibitem[{{Paxton} {et~al.}(2019){Paxton}, {Smolec}, {Schwab}, {Gautschy}, {Bildsten}, {Cantiello}, {Dotter}, {Farmer}, {Goldberg}, {Jermyn}, {Kanbur}, {Marchant}, {Thoul}, {Townsend}, {Wolf}, {Zhang}, \& {Timmes}}]{Paxton2019}
{Paxton}, B., {Smolec}, R., {Schwab}, J., {et~al.} 2019, \apjs, 243, 10

\bibitem[{{Potekhin} \& {Chabrier}(2000)}]{Potekhin2000}
{Potekhin}, A.~Y., \& {Chabrier}, G. 2000, \pre, 62, 8554

\bibitem[{{Salaris} {et~al.}(2022){Salaris}, {Cassisi}, {Pietrinferni}, \& {Hidalgo}}]{Salaris2022}
{Salaris}, M., {Cassisi}, S., {Pietrinferni}, A., \& {Hidalgo}, S. 2022, \mnras, 509, 5197

\bibitem[{{Schwab}(2021)}]{Schwab2021}
{Schwab}, J. 2021, \apj, 916, 119

\bibitem[{{Schwab} \& {Garaud}(2019)}]{SchwabGaraud2019}
{Schwab}, J., \& {Garaud}, P. 2019, \apj, 876, 10

\bibitem[{{Segretain} \& {Chabrier}(1993)}]{SegretainChabrier1993}
{Segretain}, L., \& {Chabrier}, G. 1993, \aap, 271, L13

\bibitem[{{Shen} {et~al.}(2023){Shen}, {Blouin}, \& {Breivik}}]{Shen2023}
{Shen}, K.~J., {Blouin}, S., \& {Breivik}, K. 2023, \apjl, 955, L33

\bibitem[{{Siess}(2006)}]{Siess2006}
{Siess}, L. 2006, \aap, 448, 717

\bibitem[{{Siess}(2007)}]{Siess2007}
---. 2007, \aap, 476, 893

\bibitem[{{Siess}(2010)}]{Siess2010}
---. 2010, \aap, 512, A10

\bibitem[{{Stevenson}(1980)}]{Stevenson1980}
{Stevenson}, D.~J. 1980, Journal de Physique, 41, C2 61

\bibitem[{{Tremblay} {et~al.}(2019){Tremblay}, {Fontaine}, {Gentile Fusillo}, {Dunlap}, {G{\"a}nsicke}, {Hollands}, {Hermes}, {Marsh}, {Cukanovaite}, \& {Cunningham}}]{Tremblay2019}
{Tremblay}, P.-E., {Fontaine}, G., {Gentile Fusillo}, N.~P., {et~al.} 2019, \nat, 565, 202

\bibitem[{{van Horn}(1968)}]{vanHorn1968}
{van Horn}, H.~M. 1968, \apj, 151, 227

\bibitem[{{Wu} {et~al.}(2022){Wu}, {Xiong}, \& {Wang}}]{Wu2022}
{Wu}, C., {Xiong}, H., \& {Wang}, X. 2022, \mnras, 512, 2972

\end{thebibliography}
\bibliographystyle{aasjournal}

\end{document}